\documentclass[12pt]{iopart}

\usepackage{arydshln}
\usepackage{graphicx}
\usepackage{adjustbox}
\usepackage{cite}
    
\begin{document}

\title[Parametric Analysis of a Phenomenological Constitutive Model ...]{Parametric Analysis of a Phenomenological Constitutive Model for Thermally Induced Phase Transformation in Ni-Ti Shape Memory Alloys}

\author{P. Honarmandi$^{a,*}$, A. Solomou$^b$, R. Arroyave$^a$, D. Lagoudas$^b$}

\address{$^a$ Department of Materials Science and Engineering, Texas A\&M University, College Station, Texas, USA \newline $^b$ Department of Aerospace Engineering, Texas A\&M University, College Station, Texas, USA}
\ead{hona107@tamu.edu}
\vspace{10pt}
\begin{indented}
\item[]November 2017
\end{indented}

\begin{abstract}

In this work, a thermo-mechanical model that predicts the actuation response of shape memory alloys is probabilistically calibrated against three experimental data sets simultaneously. Before calibration, a design of experiments (DOE) has been performed in order to identify the parameters most influential on the actuation response of the system and thus reduce the dimensionality of the problem. Subsequently, uncertainty quantification (UQ) of the influential parameters was carried out through Bayesian Markov Chain Monte Carlo (MCMC). The assessed uncertainties in the model parameters were then propagated to the transformation strain-temperature hysteresis curves (the model output) using first an approximate approach based on the variance-covariance matrix of the MCMC-calibrated model parameters and then an explicit propagation of uncertainty through MCMC-based sampling. Results show good agreement between model and experimental hysteresis loops after probabilistic MCMC calibration such that the experimental data are situated within 95\% Bayesian confidence intervals. The application of the MCMC-based UQ/UP approach in decision making for experimental design has also been shown by comparing the information that can be gained from running replicas around a single new experimental condition versus running experiments in different regions of the experimental space.  

\end{abstract}

\vspace{2pc}

\section{Introduction}
\label{1}

Nickle-Titanium ($Ni-Ti$) alloys are one of the most common shape memory alloys (SMAs), which are widely used in different engineering applications, such as biomedical devices and implants, microelectromechanical systems, sensors and actuators, seismic protection tools, and aerospace products and structures \cite{kang_thermomechanical_2017, kan_constitutive_2010}. Such applications are enabled by either the shape memory effect or super-elasticity, which in turn are the result of a reversible thermoelastic martensitic transformation that is triggered by applying thermal and/or mechanical loads \cite{yu_physical_2017, yu_macroscopic_2017}. 

The macroscopuc thermal actuation response of SMAs enables actuators with high specific weight ratio compared to the conventional alternatives \cite{bhaumik_nickeltitanium_2014, dilibal_nickeltitanium_2017}. Thermal actuation can occur during thermal cycles under isobaric condition where inelastic strain is recovered through the forward and reverse martensitic phase transformation induced during cooling and heating \cite{baxevanis_micromechanics_2014}. However, the forward and reverse transformations take place at different temperatures due to energy dissipation during this cyclic thermal process \cite{patoor_shape_2006}, which results in a hysteresis loop in strain-temperature space.

Under the framework of Integrated Computational Materials Engineering (ICME), completing the loop along the process-structure-property-performance chain through multiscale modeling is a necessary (albeit not sufficient) condition for the efficient and robust design of materials for specific applications. In the case of $Ni-Ti$ SMAs, different thermodynamics and kinetics modeling are usually applied to connect process and structure, such as CALculation of PHAse Diagram (CALPHAD) \cite{povoden-karadeniz_thermodynamics_2013-1, frost_thermomechanical_2010}, precipitation of secondary phases \cite{johnson_inverse_2016-1, tapia_bayesian_2017}, and phase field models~\cite{steinbach_phase-field_2011, ke_modeling_2012-1, ke_phase_2015-1}. In regard to the structure-property linkages, a number of constitutive models have been proposed over the recent decades to predict the thermo-mechanical response of $Ni-Ti$ SMAs. These models can be categorized into phenomenological~\cite{yu_macroscopic_2017-1,yu_physical_2017-1,haller_thermomechanical_2016,kan_constitutive_2010-1,jiang_propagation_2017,lagoudas_constitutive_2012,saleeb_multi-axial_2011,bodaghi_robust_2014,song_closed-form_2015,zhu_determining_2016} or micromechanics-based models~\cite{baxevanis_micromechanics_2014,mirzaeifar_micromechanical_2013,yu_micromechanical_2013, yu_micromechanical_2015,heinen_micromechanical_2012} which have been reviewed thoroughly by Patoor et al. \cite{patoor_shape_2006} and Lagoudas et al. \cite{lagoudas_shape_2006}. Generally, micro-mechanical models are high-fidelity and expensive models that use microstructural information to predict the macroscopic behavior of SMAs while phenomenological models are cheap and fast since the free energy associated with a non-microscopic homogenized material volume is taken into account for the prediction of the macroscopic responses \cite{lagoudas_constitutive_2012}.

Despite a substantial number of constitutive models for the response of SMAs, there are a very few studies associated with analysis, calibration, and uncertainty quantification (UQ) of the model parameters and subsequent uncertainty propagation (UP) from these parameters to the model responses. These uncertainties can result from different sources, including natural uncertainty (NU) from the random nature of the system, model parameter uncertainty (MPU) due to lack of knowledge and data about the model parameters, and Model structure uncertainty (MSU) as a result of incomplete or simplified physics in the model \cite{choi_inductive_2008-1}. The importance of UQ and UP, which tend to be ignored in deterministic approaches towards model parameterization, is noticeably highlighted in the case of robust design where it is sought that the system response be relatively independent of the uncertainties of its variables/parameters under the operating conditions. These inverse and forward uncertainty analyses are discussed in more detail as follows.

Model parameter estimation based on available (experimental) data for the system outputs is an inverse problem, which can be solved through deterministic or probabilistic approaches \cite{hill_mary_c._methods}. Deterministic calibration such as least squares approach yields a single best estimate based on the minimization of the squared difference between the model results and data, while probabilistic calibration provides probability distributions and uncertainty bounds for the model parameters. In different engineering fields, deterministic approaches have commonly been used to calibrate the model parameters, even though it is already widely recognized that probabilistic approaches can provide better predictions with the uncertainty/confidence bounds \cite{Kim_deterministic_2014}. 

In this regard, it is worth noting that many combinations of the model parameters usually lead to similar model outputs, but, by definition, deterministic calibration only arrives at one best estimate. Probabilistic calibration, on the other hand, provides an ensemble of possible combinations of model parameter values in the form of probability distributions \cite{larssen_forecasting_2006}.Among probabilistic approaches, Bayesian-based Markov Chain Monte Carlo algorithms have become the most commonly used tools due to the fast development in computing capabilities \cite{jackman_estimation_2000}. These sampling techniques are more robust and simple compared to the traditional analytic and numerical approaches to solve high-dimensional intractable integrals in the probabilistic calibration process \cite{olbricht_bayes_1994-1, lynch_introduction_2007-1}. In addition, MCMC methods can be used to sample from complex multivariate distributions that are usually very hard to sample using other sampling techniques such as inversion or rejection sampling \cite{lynch_introduction_2007-1}. 

MCMC sampling, however, is an expensive method, particularly in high-dimensional cases where the parameter convergence in the calibration process can be very time-consuming. A reduction in the dimensionality of the parameter space through sensitivity analysis (SA), however, can significantly improve the MCMC calibration efficiency. Sensitivity analysis (SA) is often applied to determine the impact of the parameters' variability/uncertainty on the total variability/uncertainty of the model results \cite{saltelli_variance_2010} in order to find the insensitive model parameters that can be eliminated from the calibration process. In this regard, design of experiments (DOE) is the most common approach, which requires a description of experiments (such as complete factorial design (CFD) offered by Fisher \cite{fisher_design_1966} or fractional factorial design (FFD) to consider possible combinations of pre-defined parameters' levels) besides a subsequent analysis of the experiments (such as ANalysis Of VAriance, ANOVA). 

In robust design, the final results of the model and their corresponding uncertainties are the main interest. Therefore, a UP technique should be applied by considering the trade-off between cost and information loss to propagate uncertainties from the model parameters to the model outputs \cite{ghanem_propagation_1999}. This forward problem can approximately be solved through different methods with different cost and precision, e.g., forward model analysis of the parameters' optimal combinations, statistical second moment methods, polynomial chaos expansion, etc.

In the present work, the parameter sensitivity analysis and MCMC calibration has been performed for the phenomenological constitutive model proposed by Lagoudas et al. \cite{lagoudas_constitutive_2012}. This model has been selected due to the consideration of three important features in SMAs which are usually ignored in this type of modeling: 1) the smooth transitions in thermo-mechanical responses of SMAs during forward and reverse phase transformations that take place in a range of temperatures and/or mechanical loadings, 2) the dependency of maximum transformation strain after full transformation on the magnitude of the applied stress although it does not hold at sufficiently high level of applied stresses where the material system reaches a saturated value for the maximum transformation strain resulting from the formation of all possible favored martensitic variants, and 3) the definition of a critical driving force in terms of the magnitude of the applied stress and the transformation direction \cite{lagoudas_constitutive_2012}. In section 2, this thermo-mechanical model is described thoroughly. In section 3, parameter sensitivity analysis using the combination of CFD and ANOVA are explained in details and subsequently utilized to perform the sensitivity assessment of the model parameters. In section 4, the applied MCMC sampling algorithm is described and used to probabilistically calibrate the sensitive model parameters, and then the parameter uncertainties are propagated to the model outcomes using different UP approaches. This is followed by the summary and conclusion in section 5.

\section{Description of the Constitutive Thermo-Mechanical Model}
\label{2}

In this section, the phenomenological constitutive model proposed by Lagoudas et al. \cite{lagoudas_constitutive_2012} is described in detail. In this model, a total Gibbs free energy is defined for the SMA material during the phase transformation in terms of the thermos-elastic contributions of austenite ($G^A$), martensite ($G^M$) and their interaction ($G^{mix}$) as follows:

\begin{equation}
\label{eq 1}
G(\sigma,T,\varepsilon^{t},\xi, g^{t})=(1-\xi)G^A(\sigma,T)+\xi G^M(\sigma,T)+G^{mix}(\sigma,\varepsilon^{t},g^{t})
\end{equation}

where the external state variables $\sigma$ and $T$ are the applied stress tensor and the absolute temperature, and the internal state variables $\epsilon^{t}$, $\xi$, and $g^{t}$ account for the inelastic strain produced during forward/reverse martensitic transformation, the total volume fraction of martensite, and the transformation hardening energy, respectively. Moreover, $G^A$, $G^M$, and $G^{mix}$ can be written as the following equations:

\begin{eqnarray}
\label{eq 2}
G^{A/M} &=-\frac{1}{2 \rho} \sigma : S^{A/M} \sigma - \frac{1}{\rho} \sigma : \alpha (T-T_0) \nonumber\\
& + c^{A/M} \Bigg[(T-T_0) - T \ln{\Big(\frac{T}{T_0}\Big)}\Bigg] - s_0^{A/M} T + u_0^{A/M} 
\end{eqnarray}

\begin{equation}
\label{eq 3}
G^{mix}(\sigma,\varepsilon^{t},g^{t})=-\frac{1}{\rho} \sigma : \varepsilon^{t} + \frac{1}{\rho} g^{t}
\end{equation}

The constant parameters $\rho$ and $\alpha$ are the alloy density and the thermal extension tensor, and the phase-dependent parameters $S$, $c$, $s_0$, and $u_0$ corresponds to the compliance tensor, specific heat, specific entropy, and specific internal energy, respectively. It should be noted that the notation $( . : . )$ shows the inner product of two second-order tensors.

Before applying the thermodynamics laws, an evolution relationship is defined for the transformation strain in terms of the volume fraction of martensite during forward and reverse transformations \cite{lagoudas_constitutive_2012}:

\begin{equation}
\label{eq 4}
\dot{\varepsilon}^{t}=\Lambda^{t} \dot{\xi} \ \ , \ \ \Lambda^{t}=\Bigg\{
                \begin{array}{l}
                  \Lambda_{fwd}^{t} \ \ , \ \ \dot{\xi}>0 \\
                  \Lambda_{rev}^{t} \ \ , \ \ \dot{\xi}<0
                \end{array}
\end{equation}

where $\Lambda^{t}$ represents the transformation direction tensor, which is expressed for both directions as:

\begin{equation}
\label{eq 5}
\Lambda_{fwd}^{t}=\frac{3}{2} H^{cur} \Big(\frac{\sigma^{'}}{\bar\sigma}\Big) \ \ , \ \ 
\Lambda_{rev}^{t}=\frac{\epsilon^{t}}{\xi} 
\end{equation}

$\bar\sigma=\sqrt{(3/2) \sigma^{'} : \sigma^{'}}$ is the effective stress where $\sigma^{'}$ is the deviatoric stress, and $H^{cur}$ is the maximum transformation strain after full matensitic transformation as a function of the effective stress:

\begin{equation}
\label{eq 6}
H^{cur}(\bar\sigma)=H_{sat}(1-e^{-k\bar\sigma})
\end{equation}

where $H_{sat}$ is a saturated value for the maximum transformation strain at sufficiently high value of the effective stress, and the parameter $k$ determines the exponential rate of $H^{cur}$ variation from 0 to $H_{sat}$ \cite{baxevanis_micromechanics_2014, lagoudas_constitutive_2012}.

Another evolution model can also be written between the variation rate of the transformation hardening energy and the volume fraction of martensite during forward and reverse transformation, as follows:

\begin{equation}
\label{eq 7}
\dot{g}^{t}=f^{t} \dot{\xi} \ \ , \ \ f^{t}=\Bigg\{
                \begin{array}{l}
                  f_{fwd}^{t} \ \ , \ \ \dot{\xi}>0 \\
                  f_{rev}^{t} \ \ , \ \ \dot{\xi}<0
                \end{array}
\end{equation}

$f^t$ is a direction-dependent hardening function, which has been proposed in this model such that it can capture the smooth transitions from the elastic to transformation regime and vice versa:

\begin{equation}
\label{eq 8}
\Bigg \{\begin{array}{l}
            f_{fwd}^{t}(\xi)=\frac{1}{2} a_1(1 + \xi^{n_1} - (1-\xi)^{n_2}) + a_3 \\
            f_{rev}^{t}(\xi)=\frac{1}{2} a_2(1 + \xi^{n_3} - (1-\xi)^{n_4}) - a_3
        \end{array}
\end{equation}

where $n_i$ can change in the range of (0, 1]. Closer value of $n_i$ to zero corresponds to more smooth transition at the beginning or the end of the forward phase transformation (in the case of $i=1$ or $2$, i.e., $n_1$ or $n_2$) or their counterparts for the reverse phase transformation (in the case of $i=3$ or $4$, i.e., $n_3$ or $n_4$).

Now, the first and second law of thermodynamics can be applied. At a local point in the material, the conservation of energy or first law of thermodynamics can be expressed as:

\begin{equation}
\label{eq 9}
\rho \dot{u}=\sigma : \dot{\varepsilon} - \textnormal{div}(q) + \rho r
\end{equation}

where $\dot{u}$, $q$, and $r$ denote the internal energy rate, the heat flux vector, and the internal heat generation rate, respectively. The second law of thermodynamics at this local point can also be written as the Clausius-Planck inequality \cite{paglietti_mathematical_1977}:

\begin{equation}
\label{eq 10}
\rho \dot{s} + \frac{1}{T} \textnormal{div}(q) - \frac{\rho r}{T} \geq 0
\end{equation}

It can turn into the following inequality through multiplying two sides of the inequality by T and substituting div$(q)$ by its equivalent obtained from equation \ref{eq 9}:

\begin{equation}
\label{eq 11}
\rho \dot{s} T + \sigma : \dot{\varepsilon} - \rho \dot{u} \geq 0
\end{equation}

It should be noted that Gibbs free energy can be defined in terms of internal energy using the Legendre transformation, as follows:

\begin{equation}
\label{eq 12}
G=u - \frac{1}{\rho} \sigma : \varepsilon - s T
\end{equation}

If $\dot{u}$ in equation \ref{eq 11} is substituted by the first derivative of obtained u in Equation \ref{eq 12} with respect to time, the following thermodynamic constraint is resulted for the Gibbs free energy rate:

\begin{equation}
\label{eq 13}
-\rho \dot{G} - \dot\sigma : \varepsilon - \rho s \dot{T} \geq 0
\end{equation}

Using the chain rule, the second law of thermodynamics can turn into:

\begin{equation}
\label{eq 14}
\hspace*{-1.5cm} -\rho \Big( \partial_\sigma{G} : \dot{\sigma} + \partial_T{G} : \dot{T} + \partial_{\varepsilon^{t}}{G} : \dot{\varepsilon}^{t} + \partial_\xi{G} : \dot{\xi} + \partial_{g^{t}}{G} : \dot{g}^{t} \Big) - \dot\sigma : \varepsilon - \rho s \dot{T} \geq 0
\end{equation}

Here, $-\rho \partial_\xi{G}=p$, $-\rho \partial_{\varepsilon^{t}}{G}=\sigma$, and $-\rho \partial_{g^{t}}{G}=-1$ are three generalized thermodynamic forces. In addition, the total infinitesimal strain and entropy can be expressed as follows based on Coleman and Noll approach \cite{coleman_thermodynamics_1963}:

\begin{equation}
\label{eq 15}
\varepsilon=-\rho \partial_\sigma{G}=S \sigma + \alpha (T-T_0) + \varepsilon^{t}
\end{equation}

\begin{equation}
\label{eq 16}
s=-\partial_T{G}=\frac{1}{\rho} \alpha : \sigma + c \ln{\Big( \frac{T}{T_0} \Big) + s_0}
\end{equation}

Therefore, the final inequality for the second law of thermodynamics can be expressed as:

\begin{equation}
\label{eq 17}
p \dot{\xi} + \sigma : \dot\varepsilon^{t} - \dot{g}^{t} \geq 0
\end{equation}

where the generalized thermodynamic force p can be obtained by taking derivative of total $G$ in equation \ref{eq 1} with respect to $\xi$:

\begin{eqnarray}
\label{eq 18}
p &=-\rho \partial_\xi{G}=G^M(\sigma,T)-G^A(\sigma,T) \nonumber\\
&=\frac{1}{2} \sigma : \Delta S \sigma + \sigma : \Delta\alpha (T-T_0) - \rho \Delta c \Bigg[(T-T_0) - T \ln{\Big(\frac{T}{T_0}\Big)}\Bigg] \nonumber\\
&+ \rho \Delta s_0 T - \rho \Delta u_0
\end{eqnarray}

As observed in this equation, $p$ is proportional to $ΔG=G^M-G^A$ that is the difference between the Gibbs free energy of pure martensite and pure austenite phase.

Inserting equations \ref{eq 4} and \ref{eq 7} into equation \ref{eq 17} results in:

\begin{equation}
\label{eq 19}
(\sigma : \Lambda^{t} + p - f^{t}) \dot{\xi}=\pi^{t} \dot{\xi} \geq 0
\end{equation}

where $\pi^{t}$ is the total thermodynamic force which must be positive during the forward transformation (when $\dot\xi>0$) and negative during the reverse transformation to hold the second law of thermodynamics (when $\dot\xi<0$). In this model, this thermodynamic driving force must reach a stress-dependent critical thermodynamic driving force ($Y^{t}$) in order to start and continue the phase transformation. Different values of $Y^{t}$ for forward and reverse phase transformation lead to a hyper-surface for each transformation direction, as follows: 

\begin{equation}
\label{eq 20}
\Phi^{t}(\sigma,T,\xi)=0 \ \ , \ \ \Phi^{t}(\sigma,T,\xi)=\Bigg \{\begin{array}{l}
            \Phi_{fwd}^{t}=\pi_{fwd}^{t} - Y_{fwd}^{t} \ \ , \ \ \dot\xi > 0 \\
            \Phi_{rev}^{t}=-\pi_{rev}^{t} - Y_{rev}^{t} \ \ , \ \ \dot\xi < 0
        \end{array}
\end{equation}

where,

\begin{equation}
\label{eq 21}
Y^{t}(\sigma)=\Bigg \{\begin{array}{l}
            Y_{fwd}^{t}=Y_0^{t} + D \sigma : \Lambda_{fwd}^{t} \ \ , \ \ \dot\xi > 0 \\
            Y_{rev}^{t}=Y_0^{t} + D \sigma : \Lambda_{rev}^{t} \ \ , \ \ \dot\xi < 0
        \end{array}
\end{equation}

$Y_0^{t}$ is a constant, and the model parameter $D$ indicates the dependency of the critical value for driving force on the applied stress.

In the case of iso-baric condition ($\sigma$=cons), the value of $\xi$ can be obtained at any specific $T$ during forward/reverse phase transformation using equation \ref{eq 20}. Since $\varepsilon^{t}$ is directly related to $\xi$, a hysteresis loop can be obtained in transformation strain-temperature diagram if $T$ alters from martensite start temperature ($M_s$) to martensite finish temperature ($M_f$) and from austenite start temperature ($A_s$) to austenite finish temperature ($A_f$) for forward and reverse phase transformation, respectively.

It is worth noting that the model parameters can be correlated to the material properties for the sake of the model calibration. In this regard, the material's properties $E^{A/M}$ and $\nu^{A/M}$ can represent the model parameters $S^{A/M}$ and $\alpha$, and $M_s$, $M_f$, $A_s$, $A_f$, $C^{A/M}$ can be associated with the model parameters $\rho \Delta s_0$, $\rho \Delta u_0$, $a_1$, $a_2$, $a_3$, $Y_0^{t}$, and $D$, where $E^{A/M}$ and $\nu^{A/M}$ are Young's modulus and Poisson's ratios of austenite/martensite phase, and $C^{A/M}$ is the lines slope of austenitic/martensitic transformation temperatures ($A_{s/f}/M_{s/f}$) in stress-temperature phase diagram \cite{baxevanis_micromechanics_2014, lagoudas_constitutive_2012}. 

\section{Sensitivity Analysis of the Model Parameters}
\label{3}

Before model calibration, SA is usually required to reduce the dimensionality of the parameter space with minimum possible loss of information about the model outcome in order to save calibration time and cost as a result of faster convergence of the model parameters.  In this work we carried a DOE based on a complete factorial design (CFD) coupled to ANOVA in order to identify the parameters most highly correlated to the output of the model.

\subsection{Complete Factorial Design}
\label{3-1}

CFD is an experimental design method that considers all possible combinations of the levels ($R$) defined for factors/variables/parameters ($N$) in a given system/model, which is equivalent to $R^{N}$ level-factor combinations. In the case of parameter experimental design, model response should be obtained for each level-parameter combination for the sake of sensitivity assessment. 

Based on the expert's intuition, 14 parameters have been considered as candidates for model calibration. Two plausible values have been selected for each parameter as lower and upper levels in the context of two-level CFD. These levels are considered as the applied initial values $pm 10\%$ of the range of the parameters in order to hold the following temperature constraints for their design combinations:

\begin{equation}
\label{eq 22}
M_f<M_s<A_s<A_f
\end{equation}


Therefore, $2^{14}=16384$ level-parameter combinations can be constructed during CFD whose responses are determined through the difference between the hysteresis loop obtained from running the model with each parameter combination and a reference hysteresis loop obtained using the the average value of lower and upper levels of the parameters. It is worth noting that any reference curve can be applied in so far as it is consistent for all the responses. Tschopp et al. \cite{tschopp_quantifying_2017} has evaluated various approaches to determine the similarity/difference of two images by comparing two vectors which represent the image features. In our case, these vectors were constructed based on the coordinate of spatial points on the resulting hysteresis loops:

\begin{equation}
\label{eq 23}
d_{SE}=\displaystyle\sum_{i} (X_i-Y_i)^2
\end{equation}

Now, the CFD results can be applied for SA of the given model parameters through ANOVA.

\subsection{Analysis of Variance}
\label{3-2}

A 14-way ANOVA has been performed in this work using "anovan" function in Matlab Machine Learning toolbox, where N is the number of independent Model Parameters--see Appendix A. 

ANOVA is a statistical hypothesis testing technique based on the deviation associated with each individual level of each factor/parameter/variable or each level combination of each interaction between factors/parameters/variables in the system from the overall mean of responses. The resulting response for each one of $2^{14}$ factorial design combinations has been used in the 14-way ANOVA to evaluate the sensitivity of 14 model parameters introduced in section \ref{2}. The ANOVA results associated with each parameter are shown in table \ref{Table 3}. In this table, the model parameters have been ranked based on their corresponding p-values in descending order, which indicate the parameters from the highest to the lowest sensitivity. According to the significance level 0.05, the first eight parameters have been selected for further uncertainty analysis in order to reduce the computational cost for the model calibration that will be discussed in section \ref{4-1}.

\begin{table}[t]
\caption{\label{Table 3} ANOVA table results showing the model parameters ranked based on their p-values in descending order.}
\begin{indented}
\item[]\begin{tabular}{@{}llllll}
\br
Source  & Sum sq. & d.f. & Mean sq. & F & Prob$>$F \\
\mr
$H_{sat}$ & 0.65941 & 1 & 0.65941 & 800.62 & 5.3041e-172 \\
$A_f(K)$ & 0.33278 & 1 & 0.33278 & 404.05 & 8.5195e-89 \\
$M_s(K)$ & 0.08907 & 1 & 0.08907 & 108.14 & 3.0049e-25 \\
$M_f(K)$ & 0.04402 & 1 & 0.04402 & 53.45 & 2.7735e-13 \\
$C^A(MPa/K)$ & 0.03488 & 1 & 0.03488 & 42.35 &  7.8628e-11 \\
$k(MPa^{-1})$ & 0.02462 & 1 & 0.02462 & 29.90 & 4.6223e-08 \\
$E^M(GPa)$ & 0.01029 & 1 & 0.01029 & 12.50 & 4.0821e-04 \\
$A_s(K)$ & 0.00323 & 1 & 0.00323 & 3.93 & 0.0476 \\
\hdashline
$E^A(GPa)$ & 0.00018 & 1 & 0.00018 & 0.21 & 0.6443 \\
$C^M(MPa/K)$ & 0.00008 & 1 & 0.00008 & 0.09 & 0.7612 \\
$n_1$ & 0 & 1 & 0 & 0 & 1 \\
$n_2$ & 0 & 1 & 0 & 0 & 1 \\
$n_3$ & 0 & 1 & 0 & 0 & 1 \\
$n_4$ & 0 & 1 & 0 & 0 & 1 \\
Error & 13.4819 & 16369 & 0.00082 \\
Total & 14.6805 & 16383 \\
\br
\end{tabular}
\end{indented}
\end{table}

\section{Bayesian Uncertainty Quantification and Propagation of the Model Parameters}
\label{4}

\subsection{Parameter Estimation and Uncertainty Quantification using Markov Chain Monte Carlo-Metropolis Hastings Algorithm}
\label{4-1}

In order to sample the parameter space, one can use Bayesian-based MCMC sampling approaches~ \cite{lynch_introduction_2007-1}. Among MCMC approaches, Metropolis-Hastings (M-H) sampling is one of the most common applied tools to probabilistically solve the inverse problem associated to probabilistic parameter calibration from experimental data. In the case of model calibration, this technique updates the existing prior knowledge about the parameters to their posterior information based on the available data from the model responses. 

In the present work, MCMC toolbox in Matlab has been used to probabilistically calibrate the eight sensitive model parameters identified in section \ref{3-2}. First, the prior parameter information, including the initial values ($\theta^0$), valid ranges, and probability density functions (PDFs) for the parameter are considered. Then, a new parameter vector/candidate $\theta^{cand}$ is randomly sampled from a proposal posterior PDF ($q$) which is defined as a multivariate Gaussian distribution in the toolbox with a mean value at $\theta^0$ and an arbitrary variance-covariance matrix. The candidate is accepted/rejected using a probabilistic criterion defined based on the M-H ratio that is expressed as follows:

\begin{equation}
\label{eq 37}
MH=\frac{p(\theta^{cand})p(D|\theta^{cand})}{p(\theta^0)p(D|\theta^0)} \frac{q(\theta^0|\theta^{cand})}{q(\theta^{cand}|\theta^0)}
\end{equation}

In this equation, the first ratio is called Metropolis ratio which is the ratio of the posterior probability of $\theta^{cand}$ to $\theta^0$ given data ($D$). In the context of the Bayes' theorem, the posterior probabilities are proportional to the prior probability times the likelihood---probability of the data given model results evaluated with the candidate model parameters---for each case. We considered the likelihood to be a Gaussian distribution centered at the data whose error is considered as the distribution variance. In the case of multiple data sets, it should be noted that the product of the likelihoods obtained from each data set can yield the total likelihood if data sets are assumed to be conditionally independent. 

The Metropolis ratio can be applied for the acceptance/rejection of the candidate when the proposal distribution is symmetric; however, this is not usually the case since the probability of moving from $\theta^0$ to $\theta^{cand}$ is not necessarily equal to the probability of the reverse move. For this reason, Hastings introduced the second ratio in equation \ref{eq 37} to involve the above-mentioned non-symmetric property of the proposal distribution \cite{lynch_introduction_2007-1, billera_geometric_2001}. Now, $min(MH\times100,100)$ is considered as the acceptance probability of the candidate~\cite{billera_geometric_2001}. According to this criterion, If the candidate is accepted, then $\theta^1=\theta^{cand}$; otherwise, $\theta^1=\theta^0$. The sampling of the new candidate and its acceptance/rejection process are sequentially repeated $n$ times by finding the M-H ratio using the new sampled candidate and the previous accepted parameter during each iteration. At the end of the procedure, N samples of the parameter vector ($\big\{\theta^0,...,\theta^N\big\}$) are generated to represent a multivariate posterior distribution for the model parameters. 

Before the convergence of the parameter samples, samples generated during the "burn-in period"---in which the different samples are causally correlated---must be removed before further analysis of the parameters. After the removal of the burn-in period, the mean values and the square root of diagonal elements in variance-covariance matrix of the remaining samples in the convergence region can be introduced as the plausible optimal values and uncertainties of the model parameters.

In our calibration work, initial values and their ranges have been determined based on the expert's beliefs, and an uniform (non-informative) probability distribution has been selected for each model parameter due to the absence of knowledge about the parameters' PDF. In addition, three experimental datasets for transformation strain-temperature hysteresis loop under different isobaric conditions have simultaneously been utilized together to calibrate the model. This has been performed through the calculation of the difference between various isobaric hysteresis curves obtained from each parameter sample and the corresponding experimental data using equation \ref{eq 23}. For the sake of the parameter calibration, these differences must simultaneously be compared with zero in the likelihood function where $M(\theta)$ and $D$ are the mentioned differences and zero, respectively. 

In this work, the likelihood variance $\sigma^2$ (the squared error of the distances between model and experimental hysteresis curves from zero) is set as an unknown hyper-parameter in the calibration process and updated in parallel with the other model parameters during MCMC sampling. The only difference is that an inverse-gamma distribution is defined as the non-informative prior PDF for this hyper-parameter, which yields the same form of distribution for the posterior PDF after applying the Bayes' theorem since the inverse-gamma distribution is the corresponding conjugate prior density for exponential likelihood functions. Therefore, $\sigma^2$ is sampled from an inverse gamma posterior distribution during MCMC process~ \cite{Gelman_Bayesian_2014}. In addition, it should be noted that an adaptive proposal distribution has been employed in this calibration work. In this scheme, an arbitrary positive covariance ($V_0$) is selected in the beginning of MCMC sampling, and then it is adapted in each next iteration using the covariance obtained from the ensemble of previous samples in the MCMC chain~\cite{haario_dram:_2006, haario_adaptive_2001}.

In this work, a constrained MCMC approach has also been developed in order to consider the constraints for the transformation temperatures mentioned in equation \ref{eq 22}. For this purpose, the likelihood is penalized by assigning an infinitely high value for the distance between model and experimental hysteresis curves in the cases that the constraints for the parameters are not satisfied. In these cases, the induced penalty results in a likelihood close to zero, which in turn results in an infinitesimally small acceptance probability for new model parameter candidates that violate constraints.  

\begin{figure}[htp]
\centering
\begin{minipage}{.49\textwidth}
  \centering
  \includegraphics[width=1\linewidth]{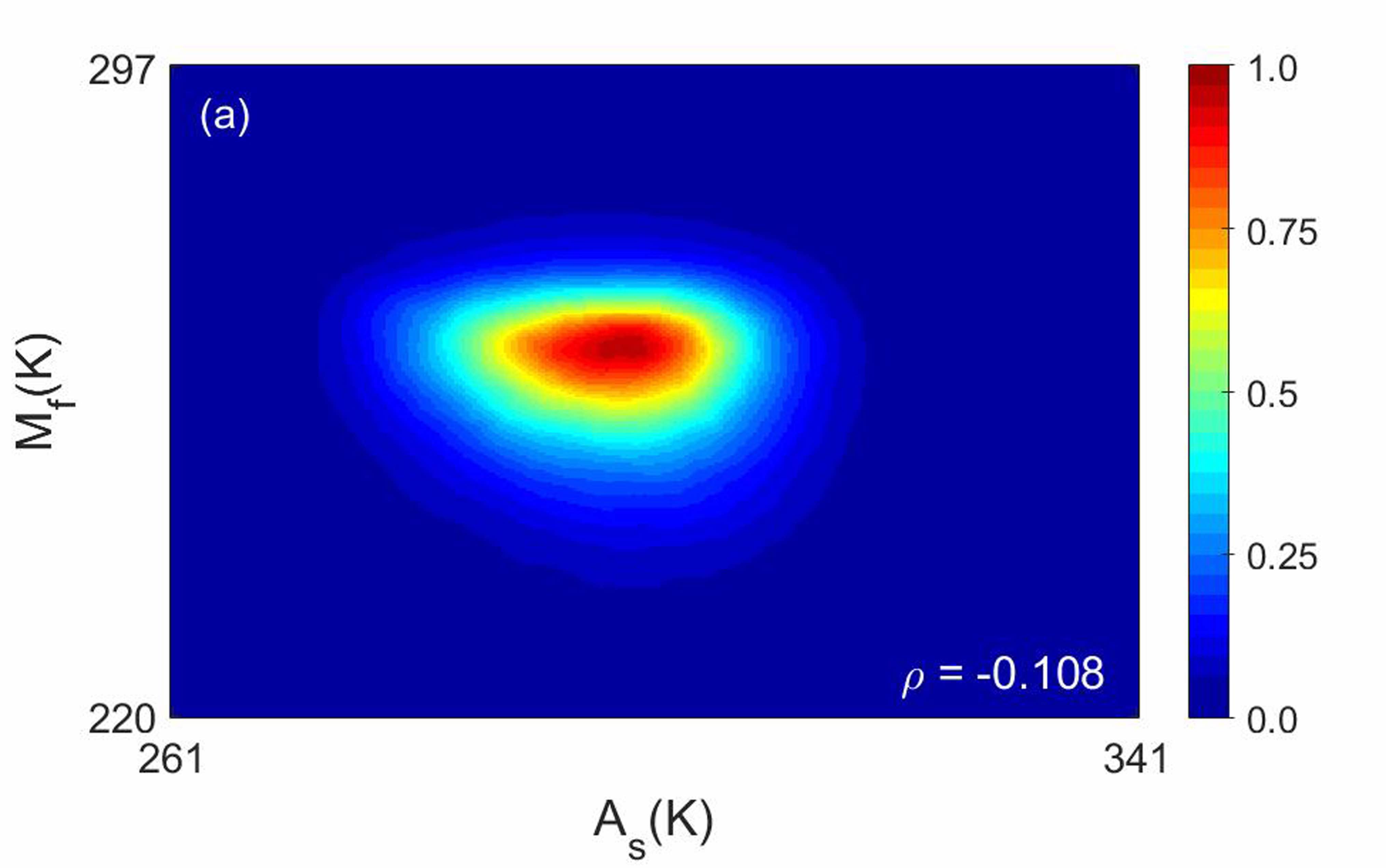}
  \label{fig 1-a}
\end{minipage}
\begin{minipage}{.49\textwidth}
  \centering
  \includegraphics[width=1\linewidth]{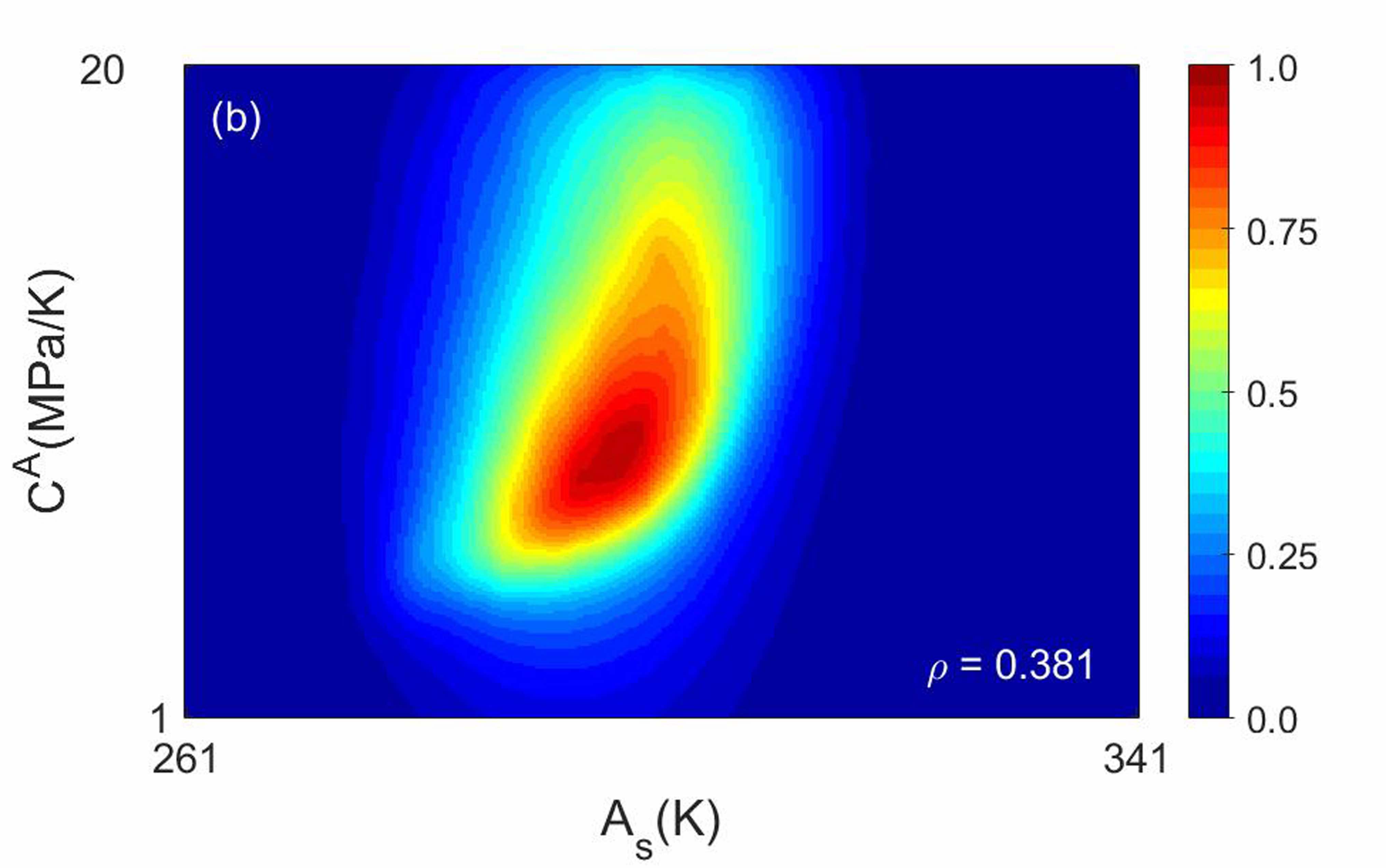}
\label{fig 1-b}
\end{minipage}
\begin{minipage}{.49\textwidth}
  \centering
  \includegraphics[width=1\linewidth]{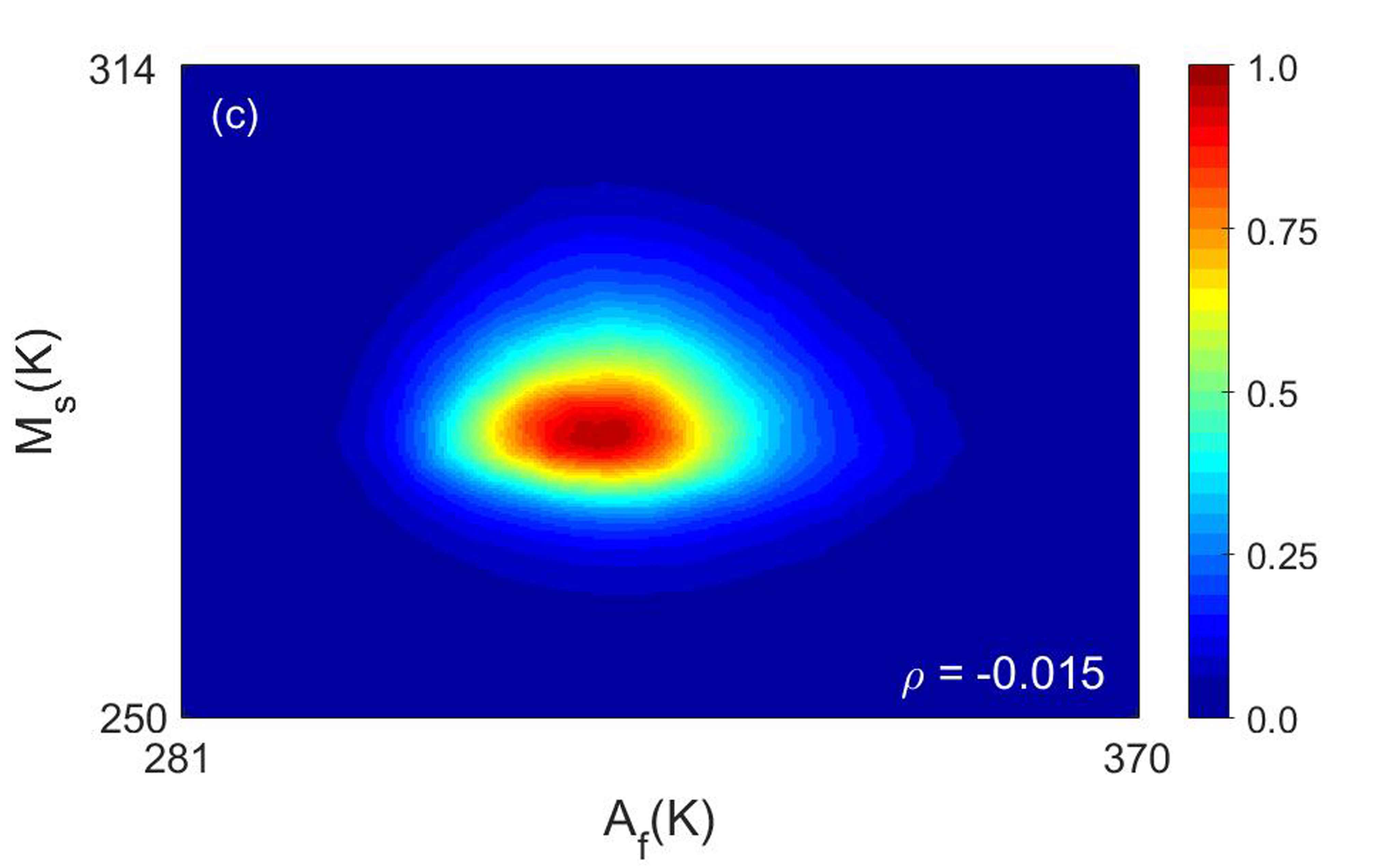}
\label{fig 1-c}
\end{minipage}
\begin{minipage}{.49\textwidth}
  \centering
  \includegraphics[width=1\linewidth]{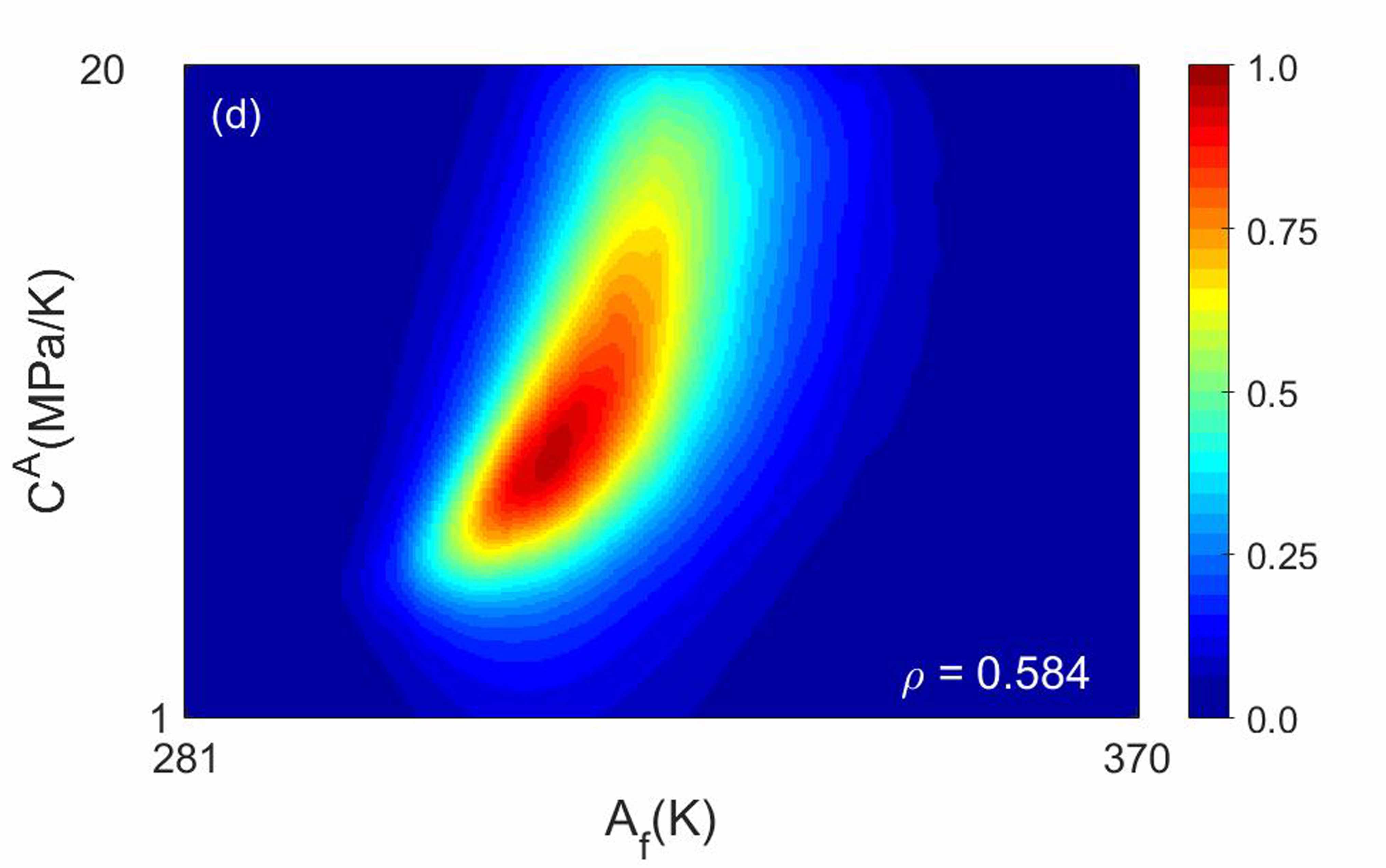}
\label{fig 1-d}
\end{minipage}
\begin{minipage}{.49\textwidth}
  \centering
  \includegraphics[width=1\linewidth]{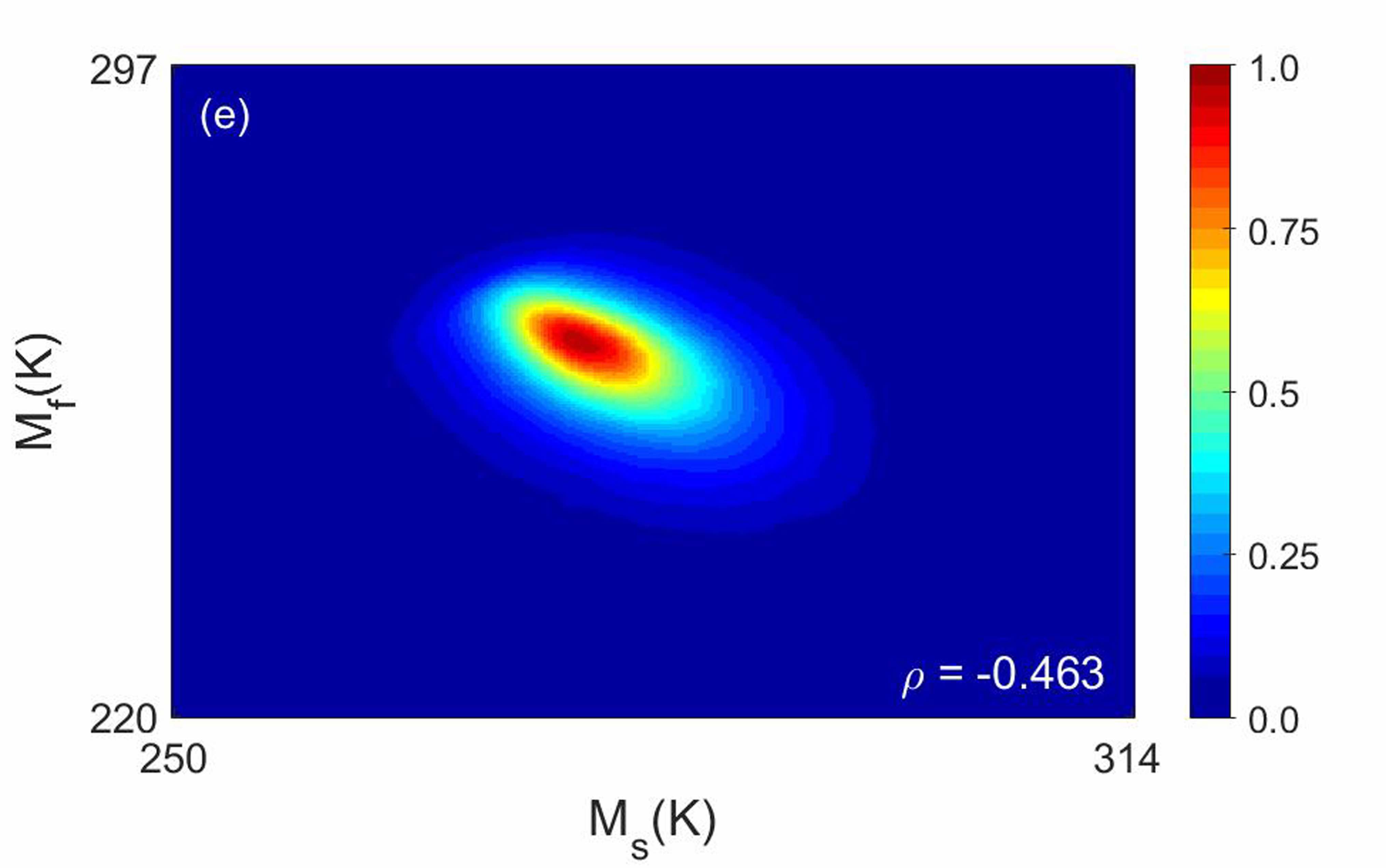}
\label{fig 1-e}
\end{minipage}
\begin{minipage}{.49\textwidth}
  \centering
  \includegraphics[width=1\linewidth]{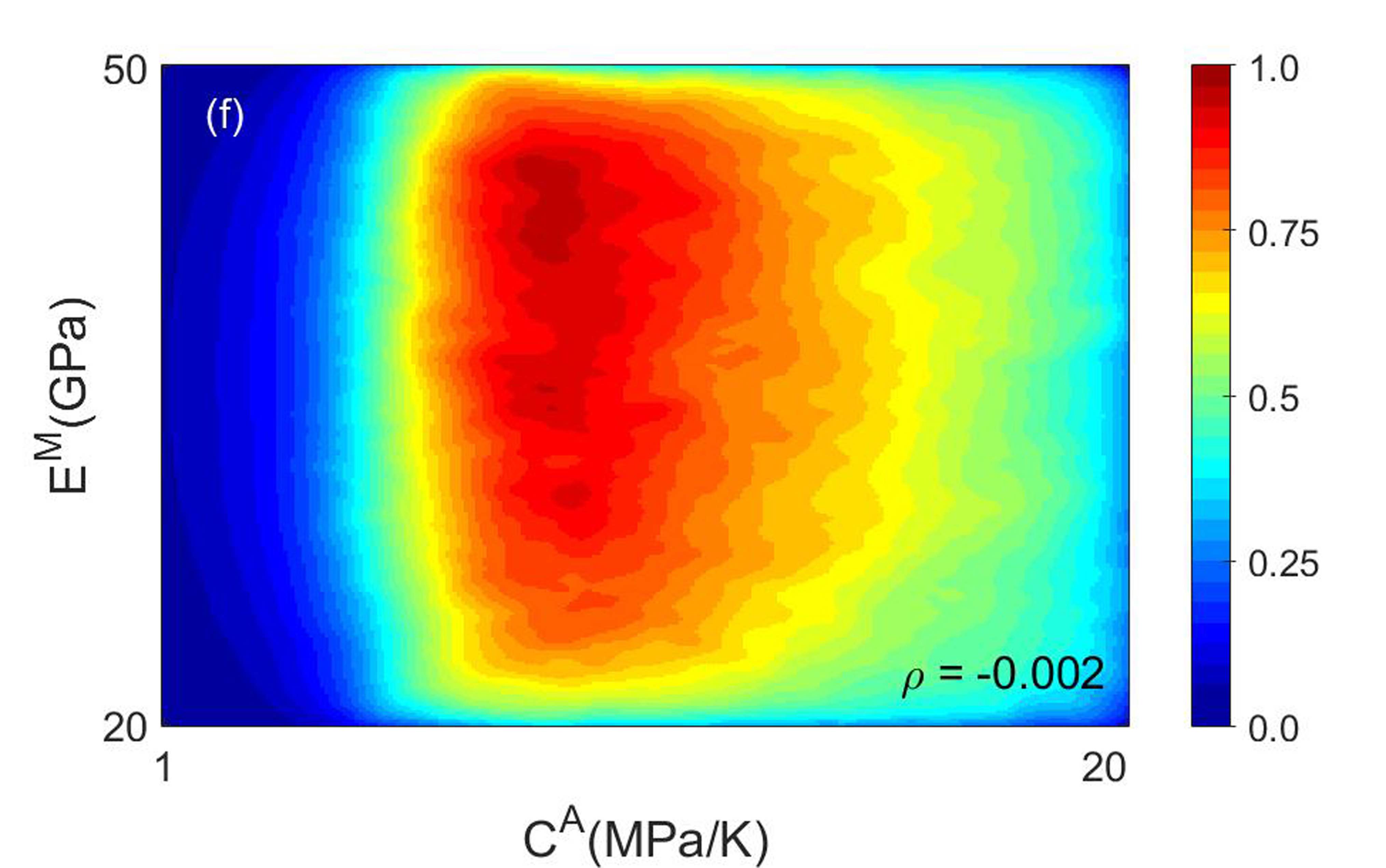}
\label{fig 1-f}
\end{minipage}
\caption{Joint frequency distributions for some of the pair model parameters in the form of 2D color graphs with normalized color bars.}
\label{fig 1}
\end{figure}

In the calibration of the sensitive model parameters, high number of samples (i.e., 200,000) are generated to ensure the convergence of the parameters and the uniqueness of the optimal response in the parameter space. The multi-variate posterior distribution after MCMC sampling can be assessed through joint and marginal frequency distributions. In figure \ref{fig 1}, some of the parameter joint distributions have been shown in 2D color graphs as examples, where the colors represents different density of samples in the parameter space, increasing from blue to red spectra. These types of graphs can also demonstrate the correlation between each pair of parameters qualitatively. For instance, negative/positive linear correlations can be inferred from the negative/positive slopes of the elliptical shapes in the graphs. The linearity of the sample distribution in the parameter space can be determined quantitatively through Pearson correlation coefficient ($-1<\rho<1$) that is expressed as:

\begin{equation}
\label{eq 38}
\rho_{X,Y}=Corr(X,Y)=\frac{cov(X,Y)}{\sigma_X \sigma_Y}
\end{equation}

where $\sigma_X$ and $\sigma_Y$ are the standard deviations for parameters $X$ and $Y$, respectively, which equal the square root of the corresponding diagonal elements in $8\times8$ variance-covariance matrix. $cov(X,Y)$ is also the covariance of $X$ and $Y$ obtained from the corresponding non-diagonal element in the variance-covariance matrix. Generally, a closer value of $\rho$ to either -1 or 1 implies a higher linear correlation between $X$ and $Y$, whereas a closer value to zero suggests their lower linear correlation. The negative or positive sign refers to the negative or positive linear correlation. The linear coefficients for all pairs of parameters in the thermo-mechanical model have been listed in table \ref{Table 4}. As can be observed in this table, most of the pair parameters are linearly uncorrelated although 6 out of 28 parameter pairs show some weak or moderate correlations with $|\rho|$ greater than 0.2. In addition, the parameters $C^A$ and $A_f$ (\ref{fig 1}-d) and the parameters $E^M$ and $C^A$ (\ref{fig 1}-f) show the highest and the lowest linear correlation, respectively, regardless of the correlation signs. 

\begin{table}[t]
\caption{\label{Table 4} Linear correlation coefficient for each pair of model parameters.}
\begin{adjustbox}{max width=\textwidth}
\begin{tabular}{@{}lllllllll}
\br
{} & $A_s(K)$ & $A_f(K)$ & $M_s(K)$ & $M_f(K)$ & $C^A(MPa/K)$ & $E^M(GPa)$ & $H_{sat}$ & $k(MPa^{-1})$ \\
\mr
$A_s(K)$ & 1 & -0.0721 & 0.2948 & -0.1080 & 0.3813 & -0.0167 & -0.1330 & -0.0378 \\
$A_f(K)$ & -0.0721 & 1 & -0.0150 & -0.0822 & 0.5836 & -0.0553 & 0.0249 & -0.0672 \\
$M_s(K)$ & 0.2948 & -0.0150 & 1 & -0.4628 & 0.1156 & -0.0547 & -0.0465 & -0.0415 \\
$M_f(K)$ & -0.1080 & -0.0822 & -0.4628 & 1 & -0.1280 & -0.0289 & -0.1355 & 0.0160 \\
$C^A(MPa/K)$ & 0.3813 & 0.5836 & 0.1156 & -0.1280 & 1 & -0.0015 & 0.0476 & -0.0405 \\
$E^M(GPa)$ & -0.0167 & -0.0553 & -0.0547 & -0.0289 & -0.0015 & 1 & 0.2771 & 0.0091 \\
$H_{sat}$ & -0.1330 & 0.0249 & -0.0465 & -0.1355 & 0.0476 & 0.2771 & 1 & -0.2103 \\
$k(MPa^{-1})$ & -0.0378 & -0.0672 & -0.0415 & 0.0160 & -0.0405 & 0.0091 & -0.2103 & 1 \\
\br
\end{tabular}
\end{adjustbox}
\end{table}

\begin{figure}[htp]
\centering
\begin{minipage}{.49\textwidth}
  \centering
  \includegraphics[width=1\linewidth]{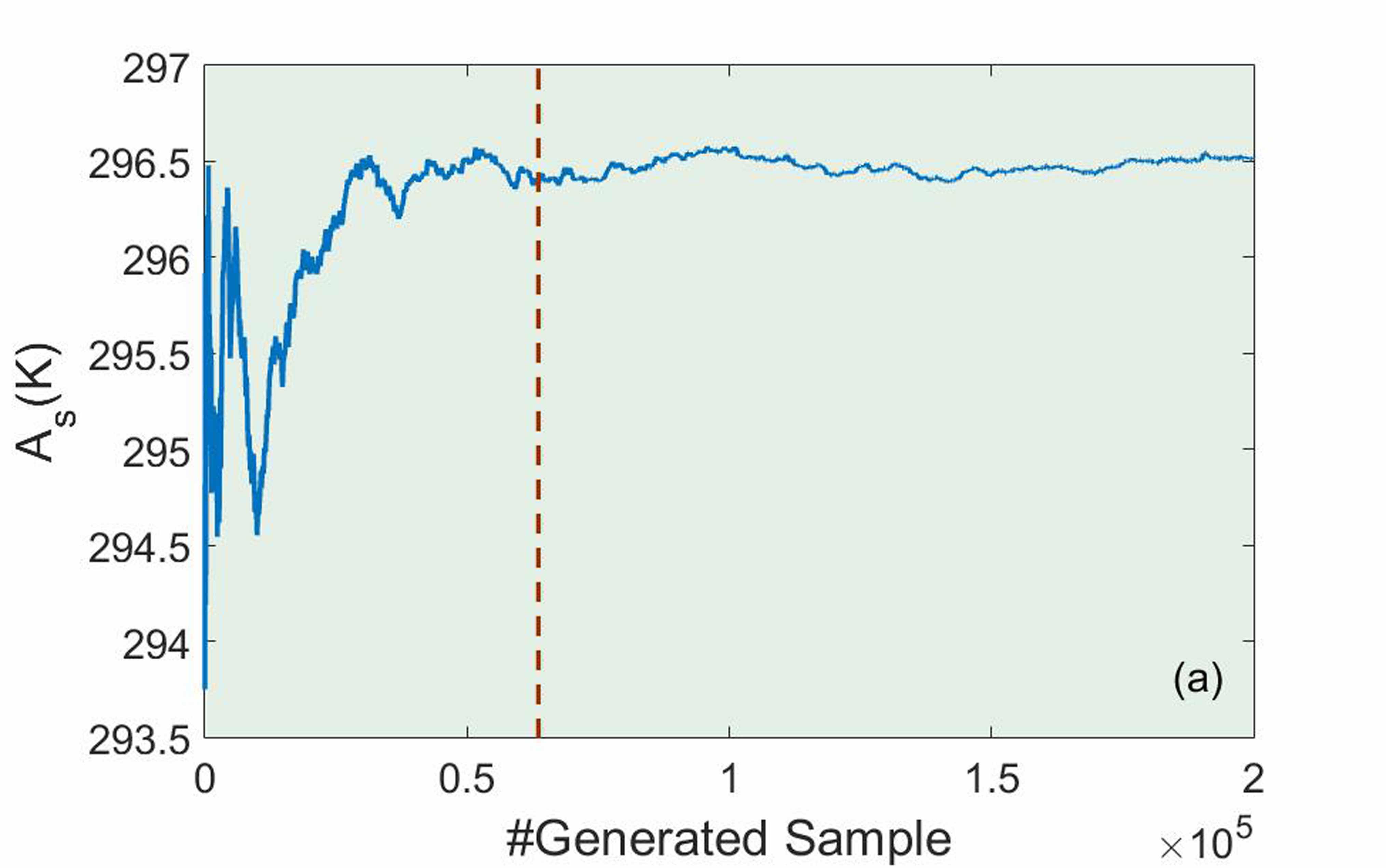}
  \label{fig 2-a}
\end{minipage}
\begin{minipage}{.49\textwidth}
  \centering
  \includegraphics[width=1\linewidth]{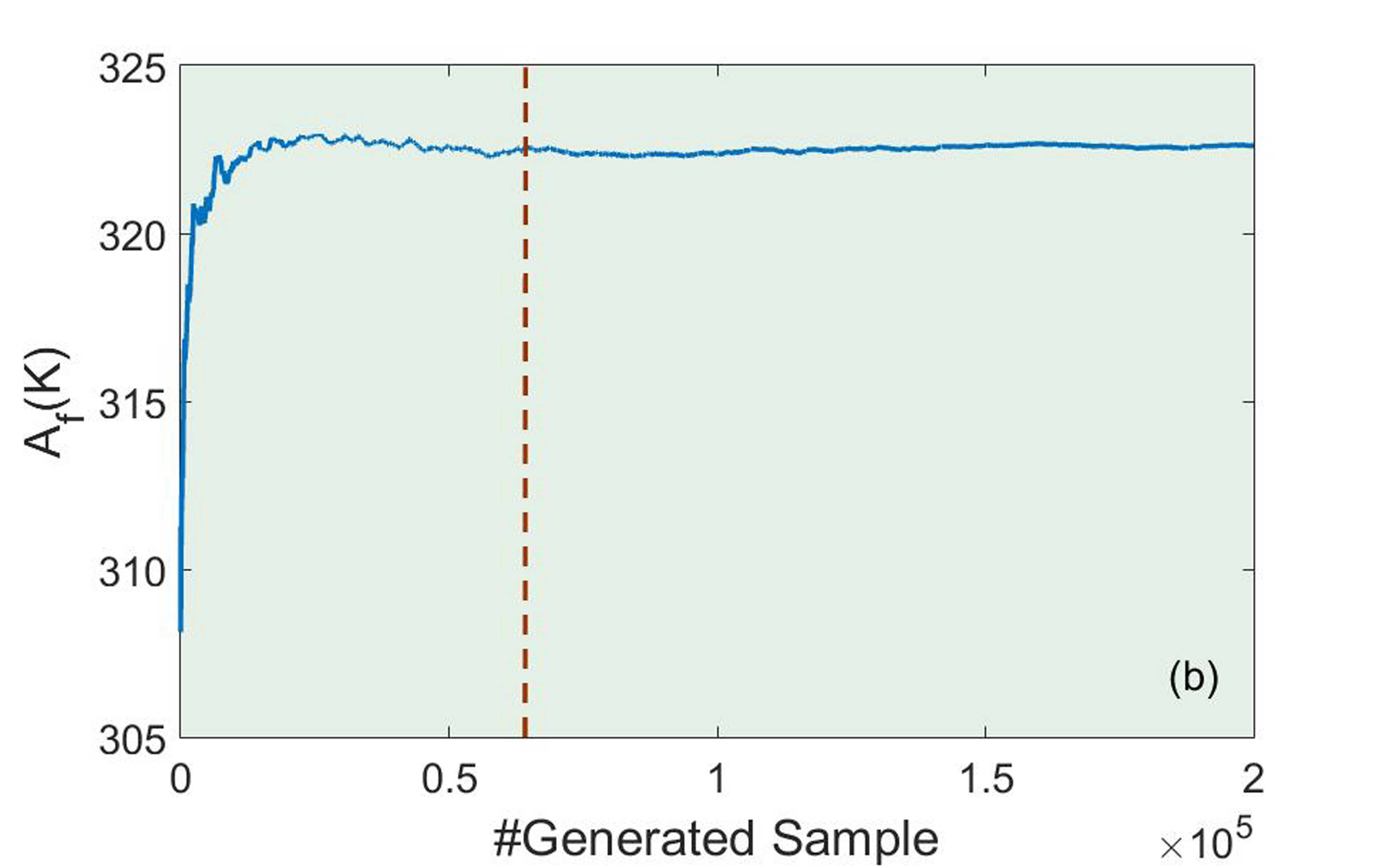}
\label{fig 2-b}
\end{minipage}
\begin{minipage}{.49\textwidth}
  \centering
  \includegraphics[width=1\linewidth]{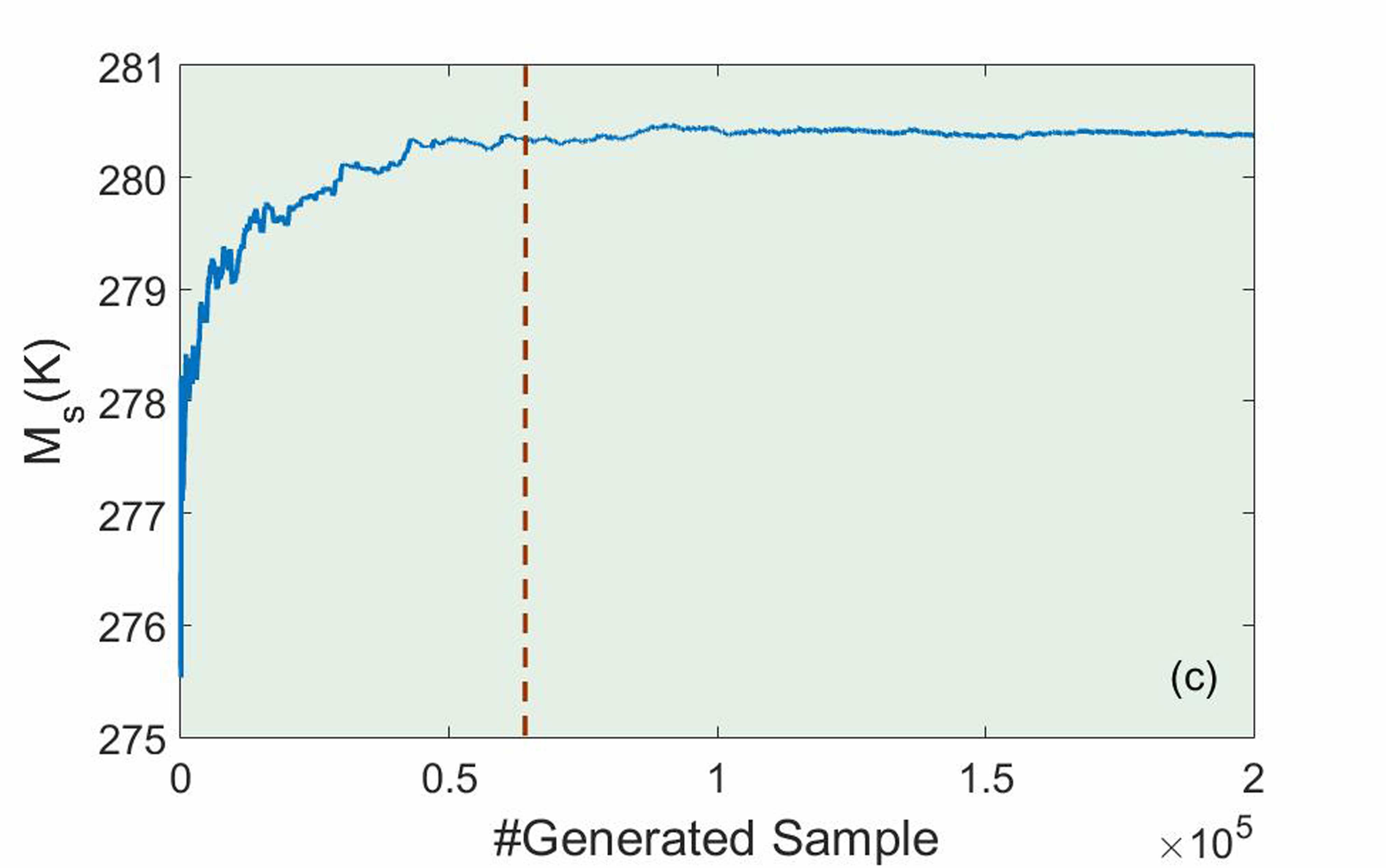}
\label{fig 2-c}
\end{minipage}
\begin{minipage}{.49\textwidth}
  \centering
  \includegraphics[width=1\linewidth]{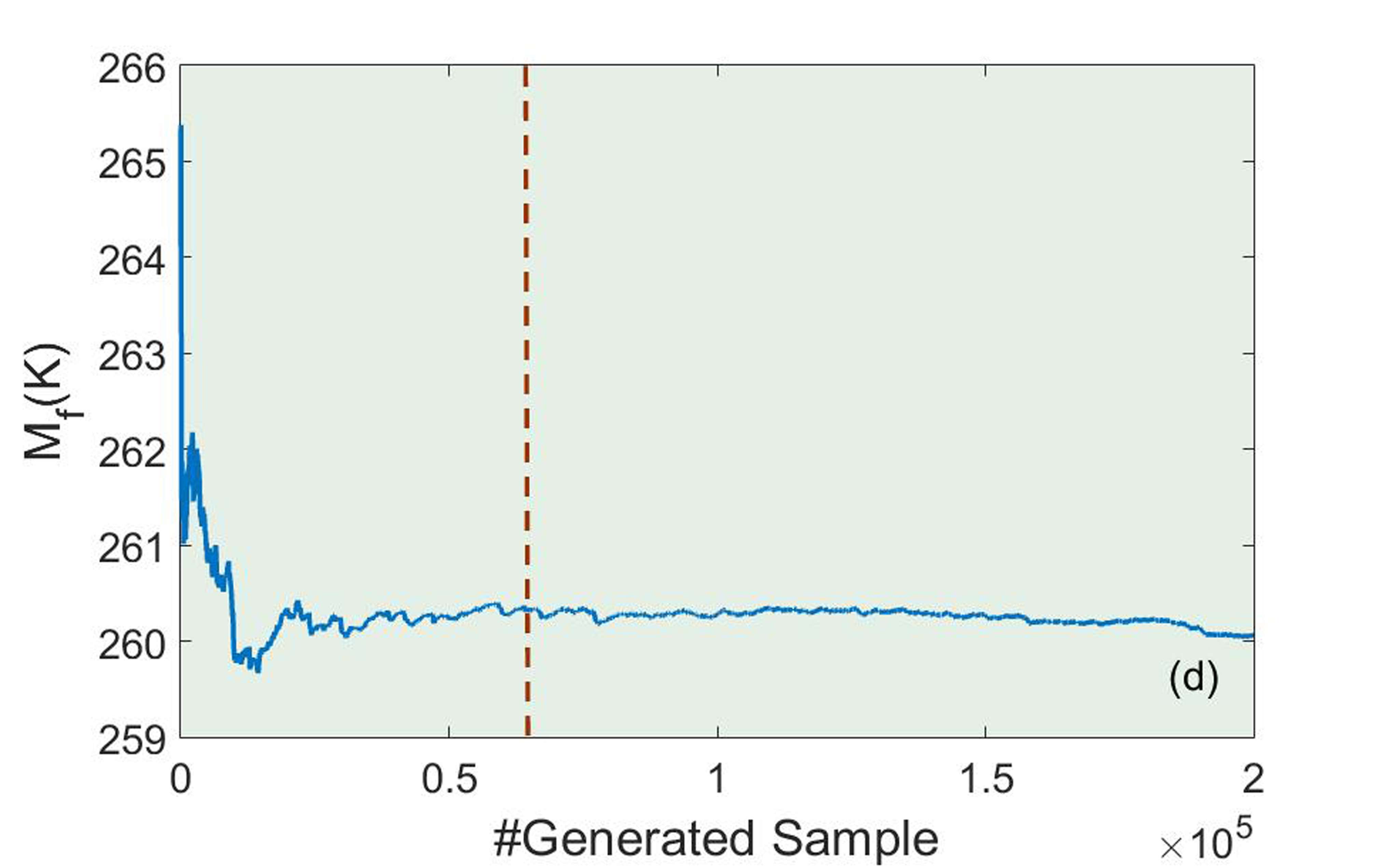}
\label{fig 2-d}
\end{minipage}
\begin{minipage}{.49\textwidth}
  \centering
  \includegraphics[width=1\linewidth]{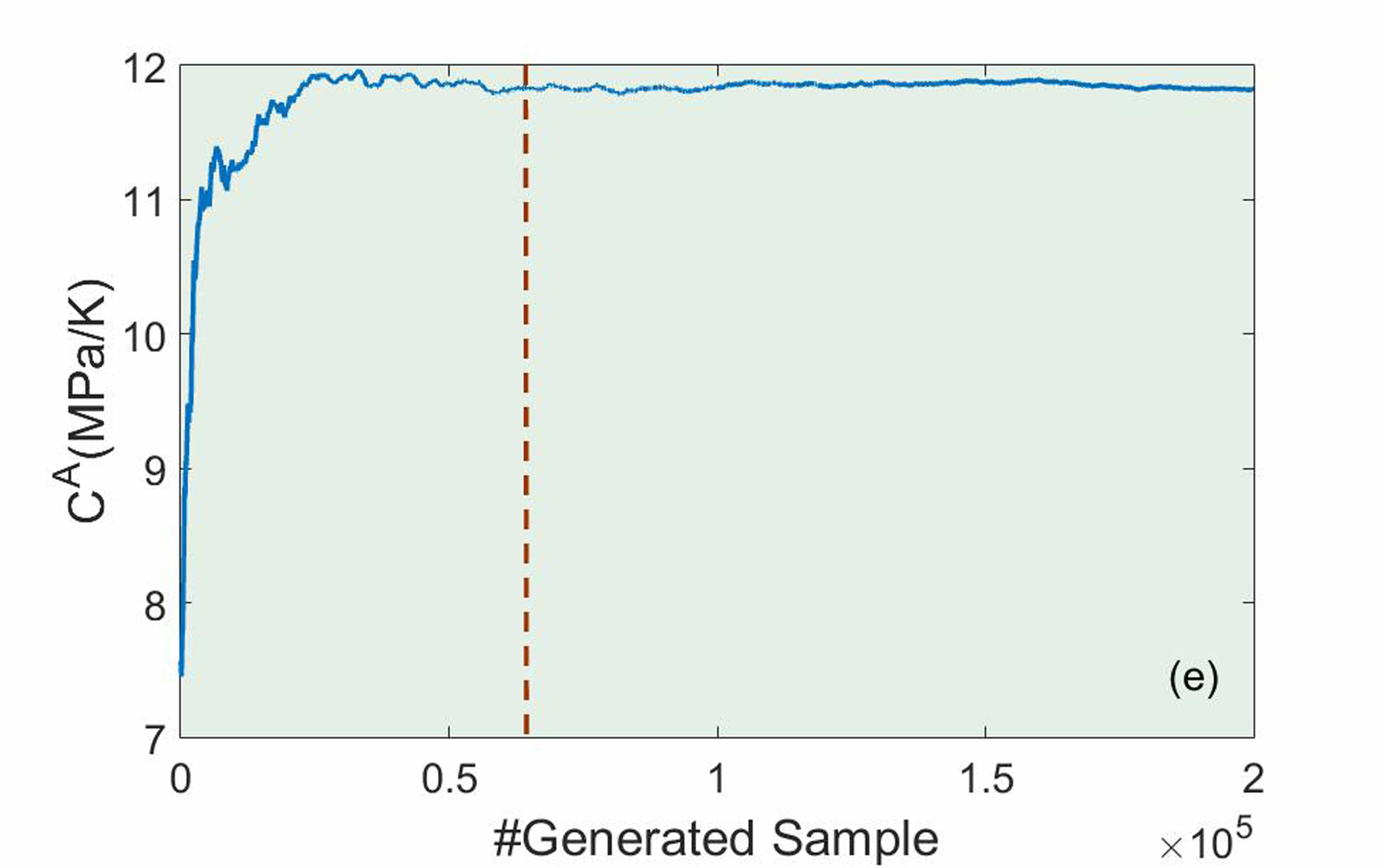}
\label{fig 2-e}
\end{minipage}
\begin{minipage}{.49\textwidth}
  \centering
  \includegraphics[width=1\linewidth]{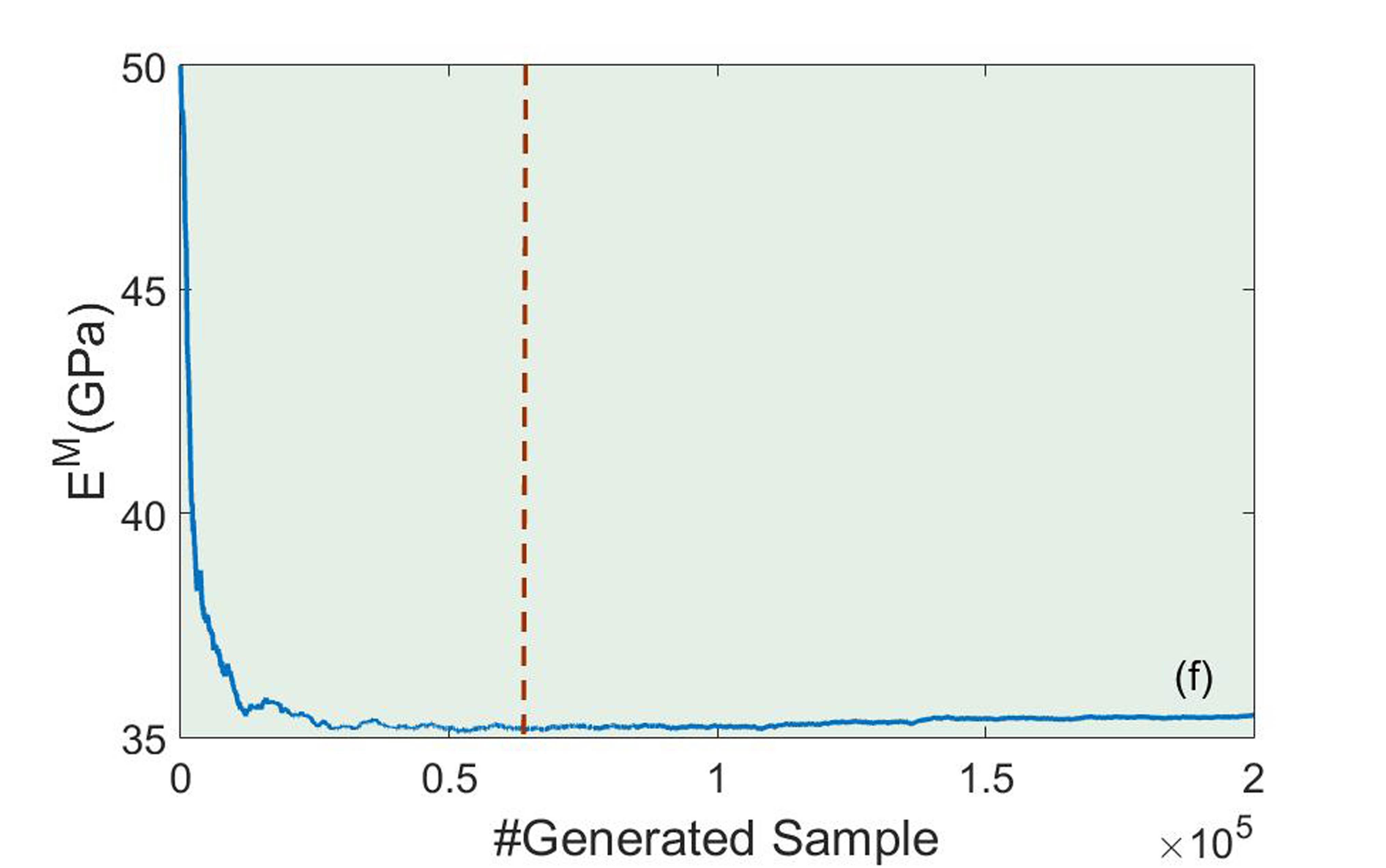}
\label{fig 2-f}
\end{minipage}
\begin{minipage}{.49\textwidth}
  \centering
  \includegraphics[width=1\linewidth]{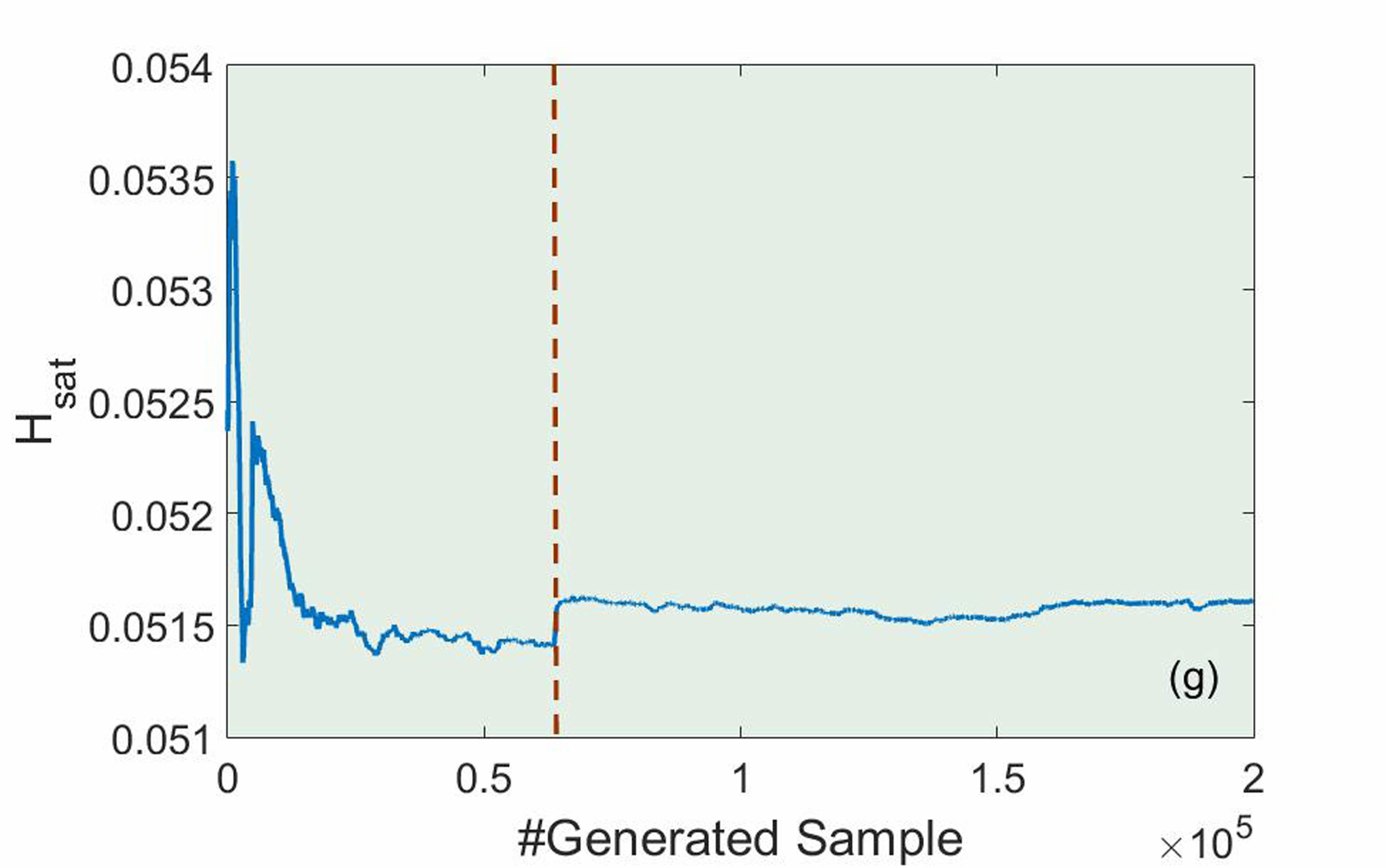}
\label{fig 2-g}
\end{minipage}
\begin{minipage}{.49\textwidth}
  \centering
  \includegraphics[width=1\linewidth]{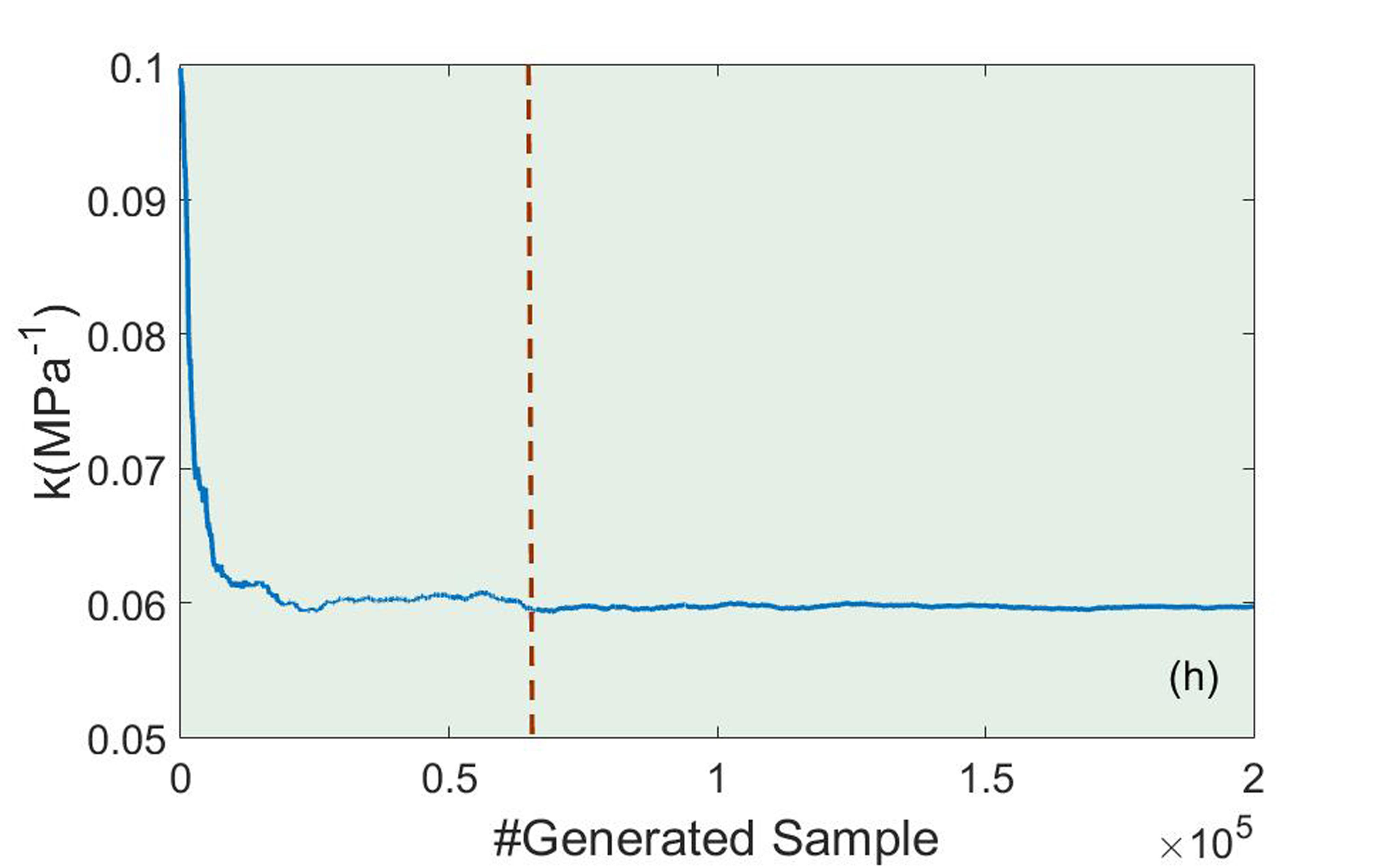}
\label{fig 2-h}
\end{minipage}
\caption{Cumulative mean distribution plot for each model parameter after MCMC sampling.}
\label{fig 2}
\end{figure}

As mentioned above, the burn-in period at the beginning of MCMC sampling must be removed before the calibration and UQ of the model parameters, which can be found using the cumulative mean distributions for the parameters, as shown in figure \ref{fig 2}. In these figures, there are some noisy trends at the beginning of MCMC sampling for the values of all the parameters, but they reach almost plateaus after about 63,000 generations that show the convergence values (mean values) for the parameters. Therefore, the first 63,000 samples are extracted from the total MCMC sample collection before further analysis. The mean and the square root of the diagonal elements in the variance-covariance matrix of the remaining samples in the convergence region can be introduced as the plausible optimal values and the uncertainty of the model parameters, which are listed in table \ref{Table 5}.

\begin{table}[t]
\caption{\label{Table 5} Plausible optimal values and uncertainties (standard deviations) of the model parameters after MCMC calibration besides their initial values.}
\begin{adjustbox}{max width=\textwidth}
\begin{tabular}{@{}lllllllll}
\br
{} & $A_s(K)$ & $A_f(K)$ & $M_s(K)$ & $M_f(K)$ & $C^A(MPa/K)$ & $E^M(GPa)$ & $H_{sat}$ & $k(MPa^{-1})$ \\
\mr
Initial Values & 307.0 & 318.0 & 300.0 & 270.0 & 9.0 & 40.0 & 0.034 & 0.02 \\
After MCMC Calibration & 296.6$\pm$8.5 & 322.6$\pm$11.3 & 280.4$\pm$6.6
 & 259.9$\pm$8.5 & 11.8$\pm$4.1 & 35.6$\pm$8.6 & 0.0517$\pm$0.0044 & 0.0595$\pm$0.0237 \\
\br
\end{tabular}
\end{adjustbox}
\end{table}

\begin{figure}[htp]
\centering
\begin{minipage}{.49\textwidth}
  \centering
  \includegraphics[width=1\linewidth]{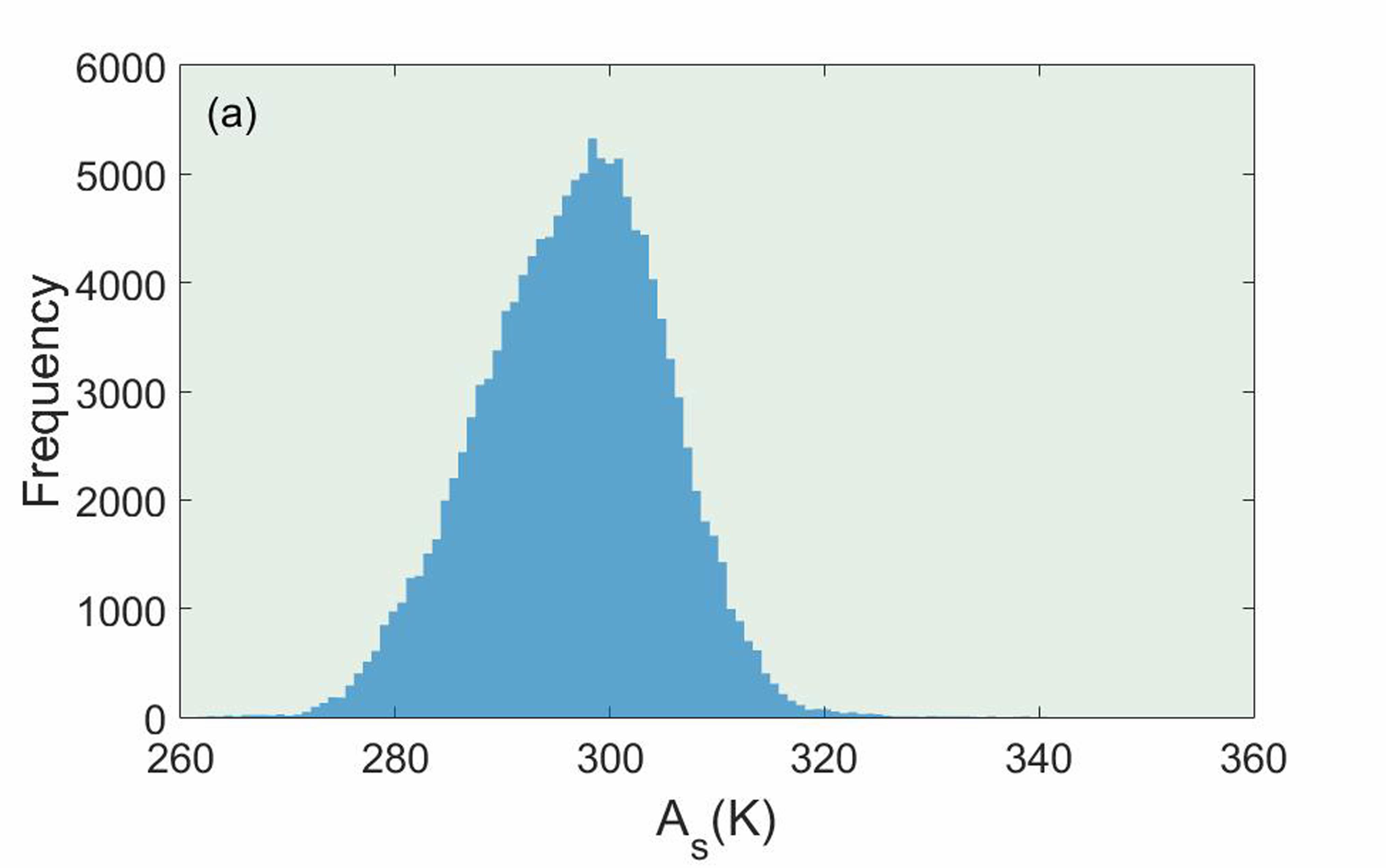}
  \label{fig 3-a}
\end{minipage}
\begin{minipage}{.49\textwidth}
  \centering
  \includegraphics[width=1\linewidth]{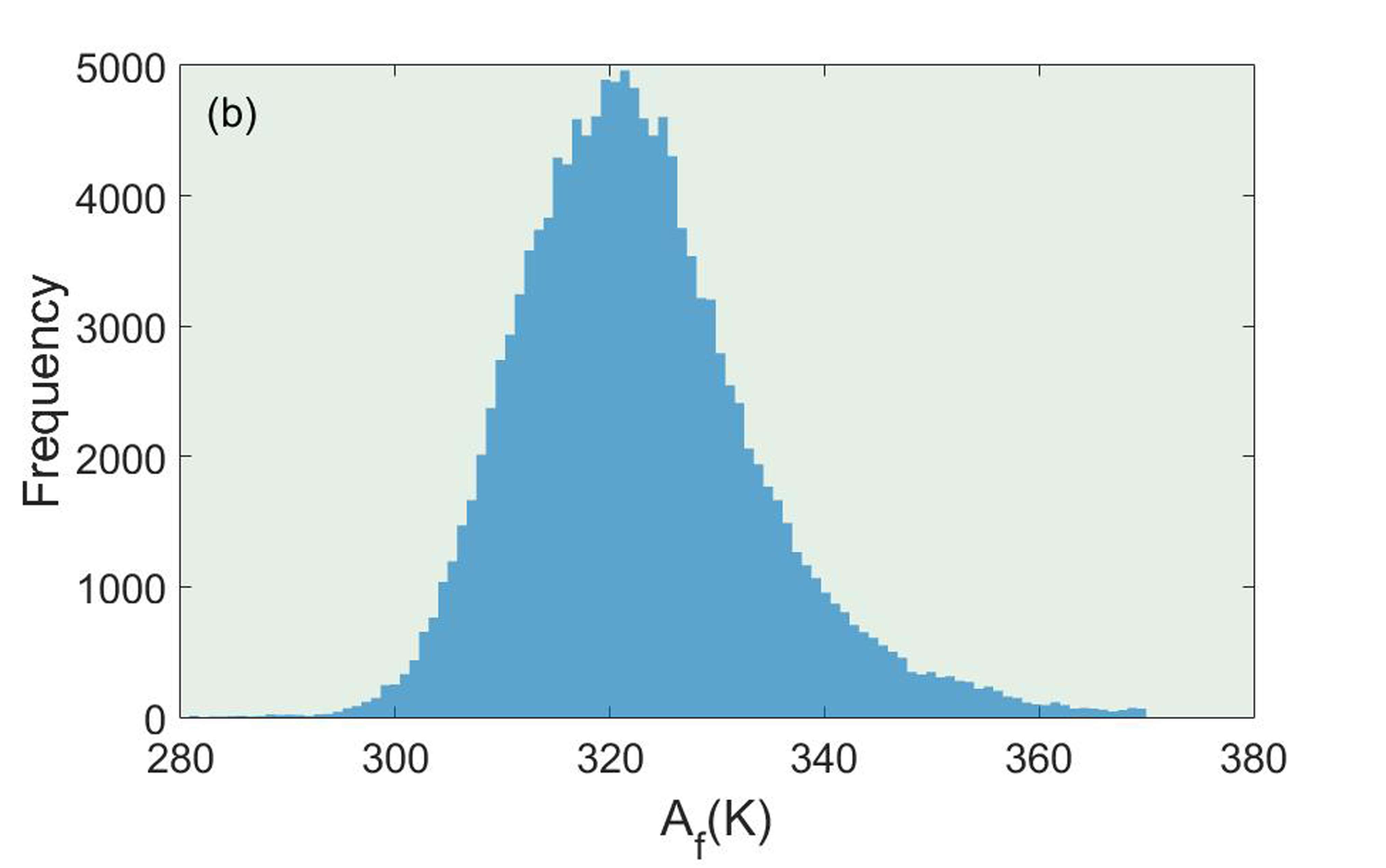}
\label{fig 3-b}
\end{minipage}
\begin{minipage}{.49\textwidth}
  \centering
  \includegraphics[width=1\linewidth]{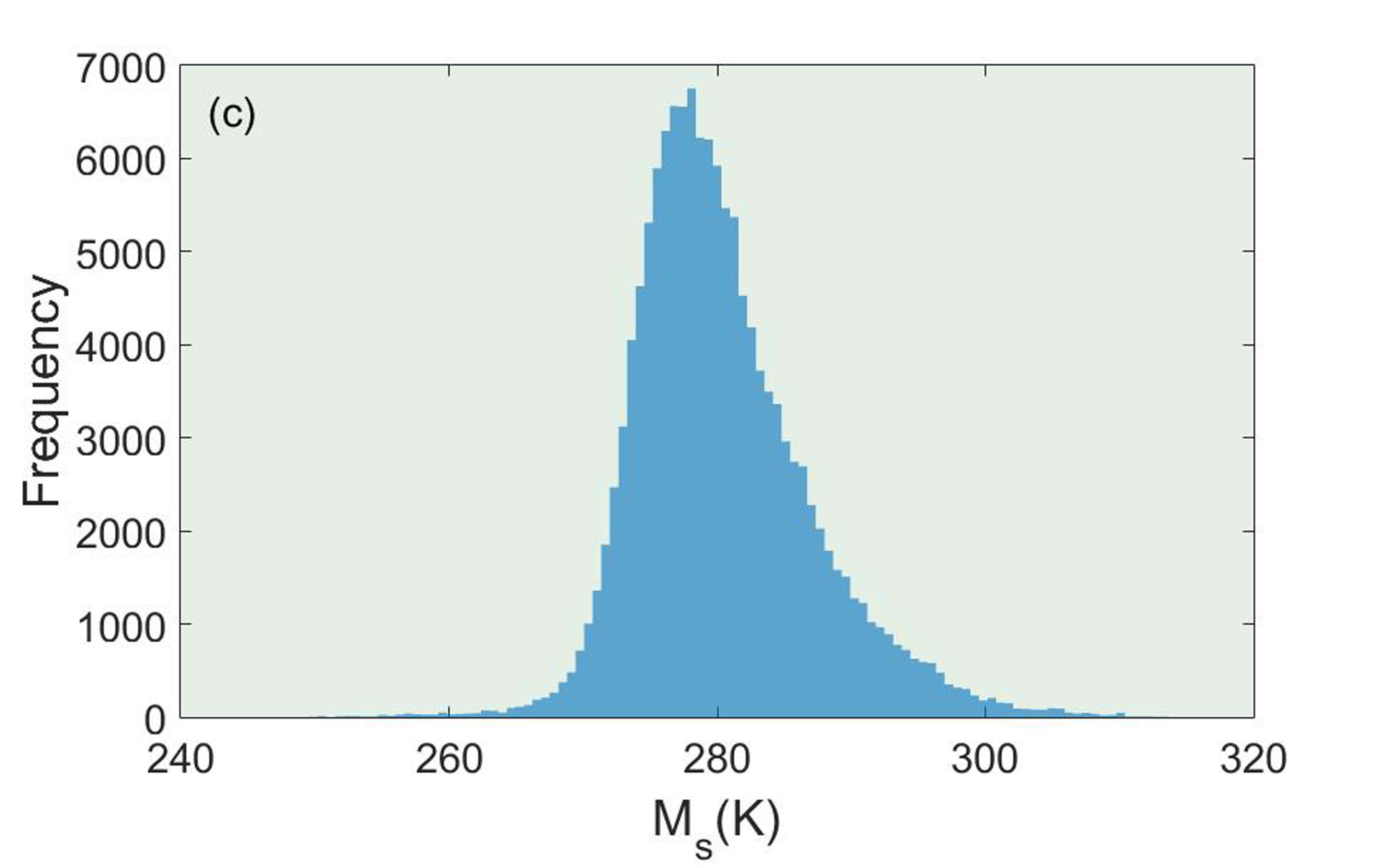}
\label{fig 3-c}
\end{minipage}
\begin{minipage}{.49\textwidth}
  \centering
  \includegraphics[width=1\linewidth]{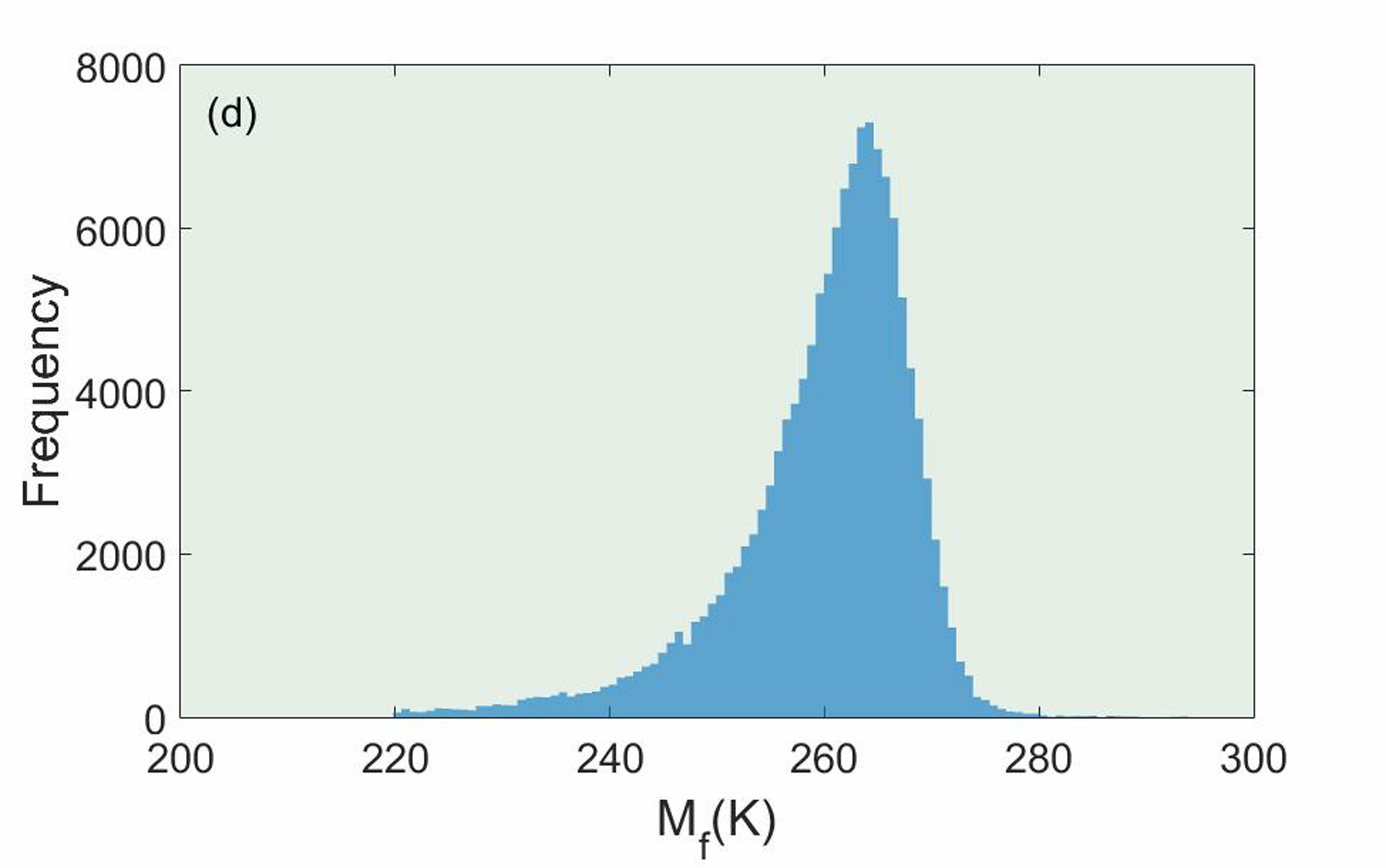}
\label{fig 3-d}
\end{minipage}
\begin{minipage}{.49\textwidth}
  \centering
  \includegraphics[width=1\linewidth]{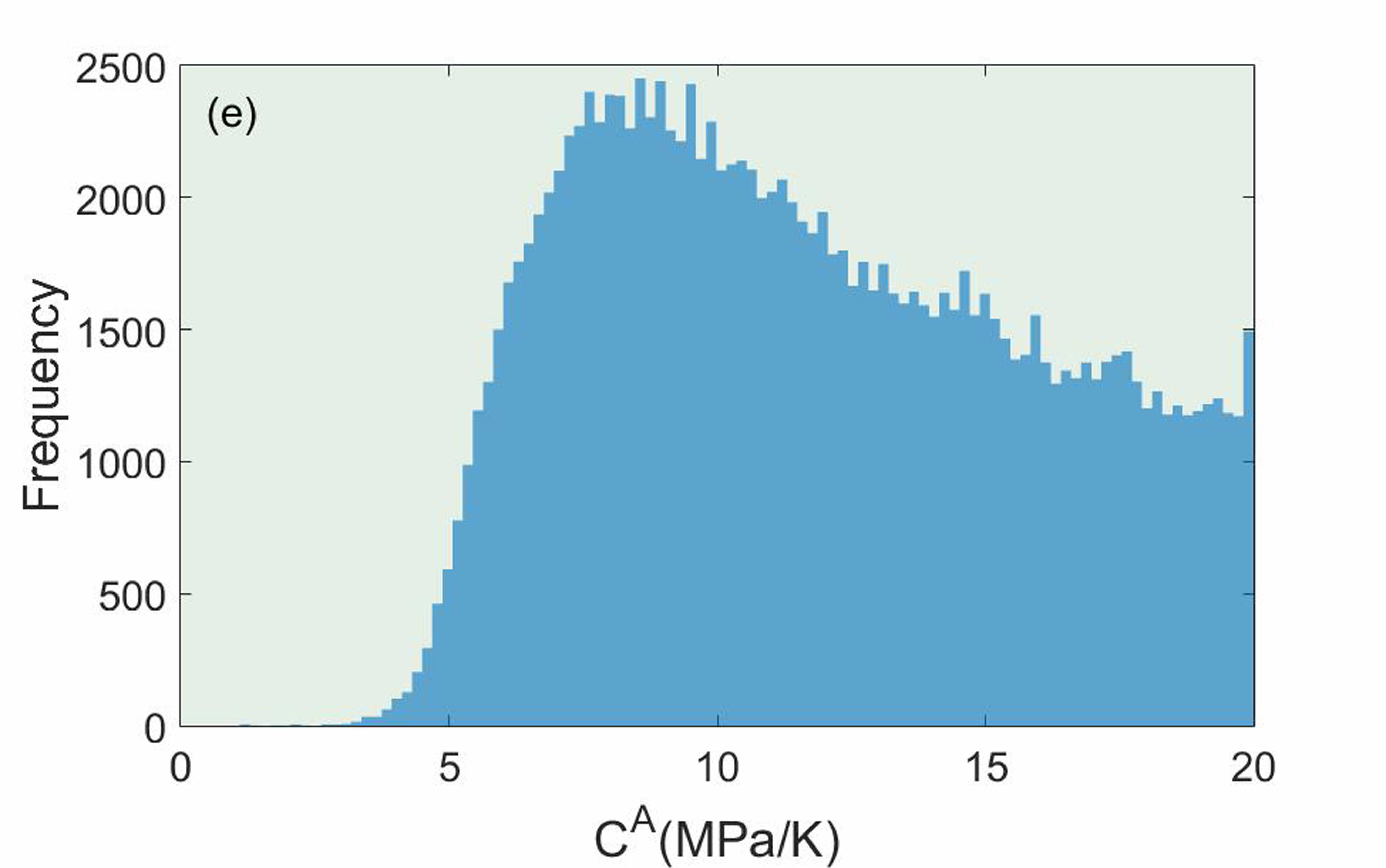}
\label{fig 3-e}
\end{minipage}
\begin{minipage}{.49\textwidth}
  \centering
  \includegraphics[width=1\linewidth]{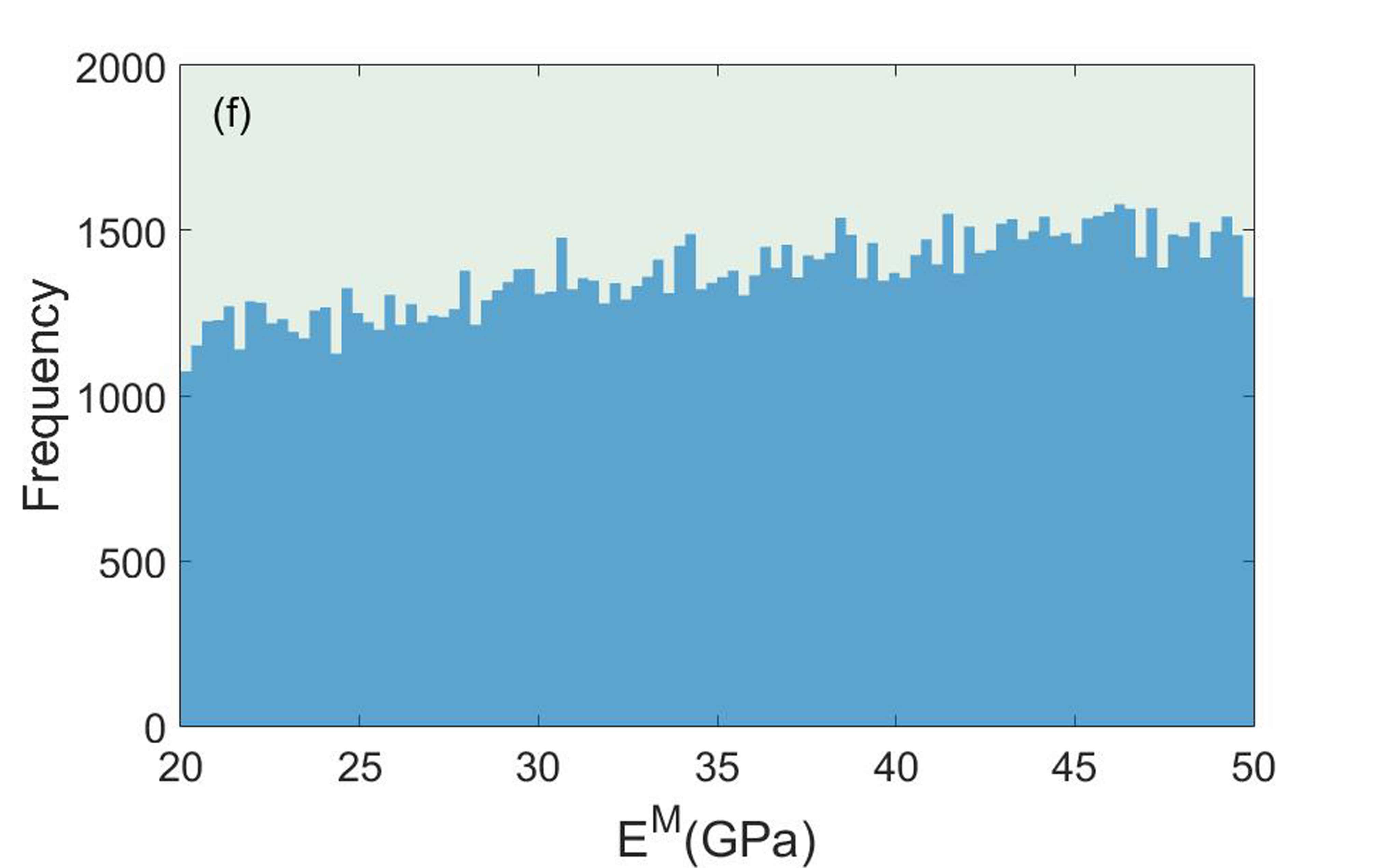}
\label{fig 3-f}
\end{minipage}
\begin{minipage}{.49\textwidth}
  \centering
  \includegraphics[width=1\linewidth]{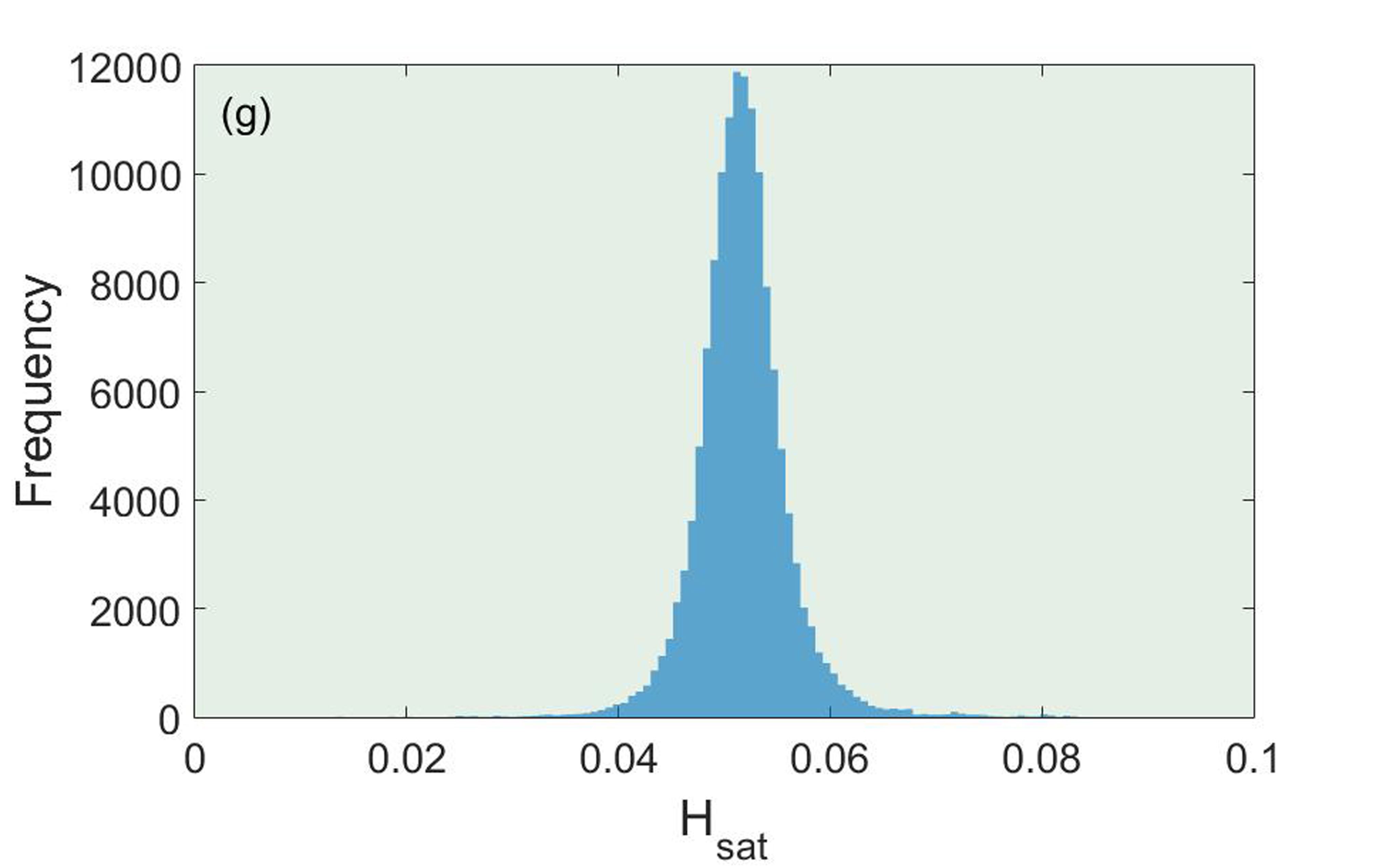}
\label{fig 3-g}
\end{minipage}
\begin{minipage}{.49\textwidth}
  \centering
  \includegraphics[width=1\linewidth]{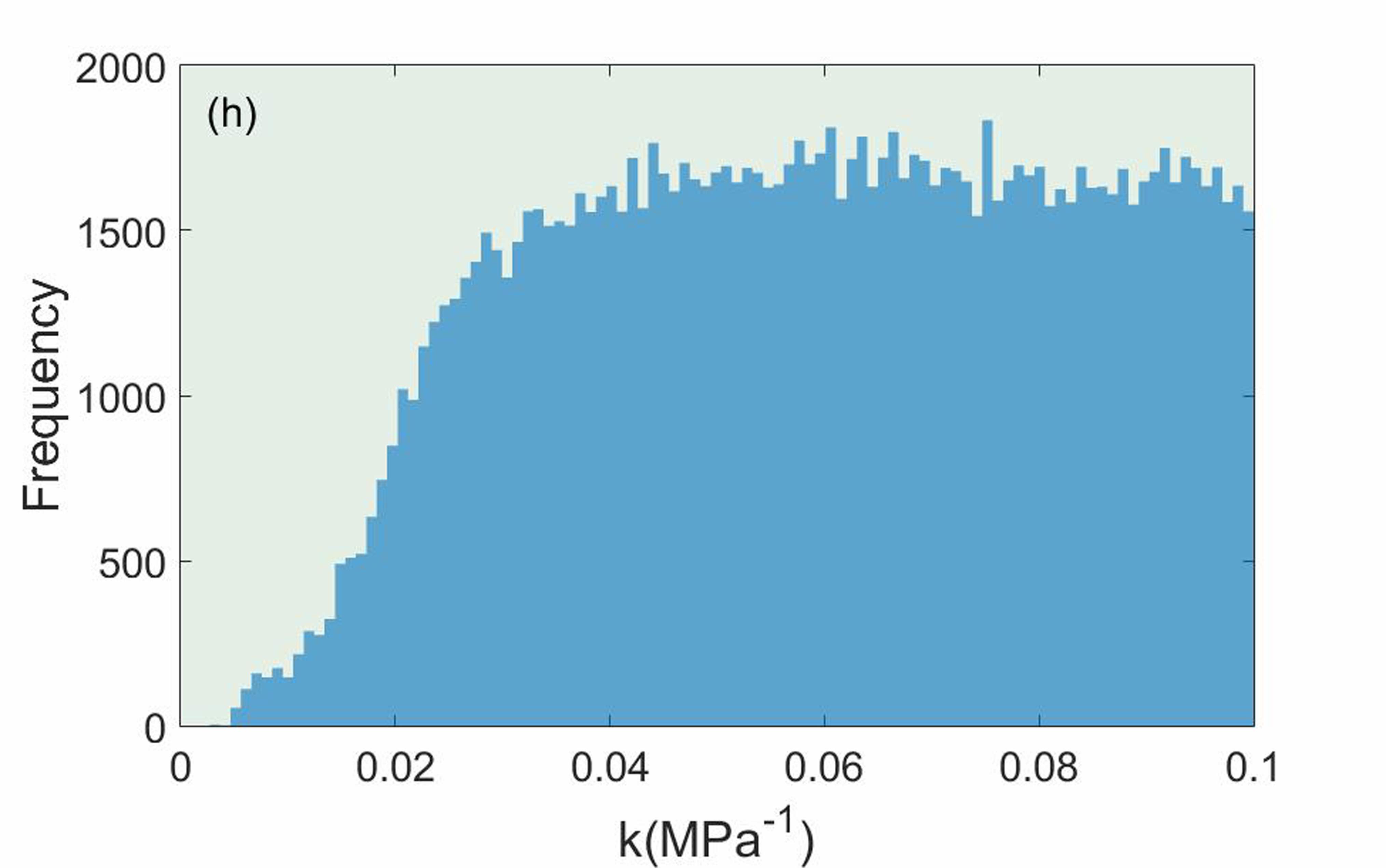}
\label{fig 3-h}
\end{minipage}
\caption{marginal posterior frequency distribution plot for each model parameter after the removal of the burn-in period from MCMC samples.}
\label{fig 3}
\end{figure}

After removal of the burn-in period, the marginal posterior frequency distribution of the parameters can be plotted using the remaining MCMC samples as observed in figure \ref{fig 3}. These plots suggest skewed Gaussian distributions for the marginal posterior frequency distribution (although it seems to be a Gaussian distribution in the case of the parameter $H_{sat}$), which are truncated from both sides by the lower and upper bounds of the parameters. However, the parameters $E^M$ and $k$ show flatter distributions due to their low sensitivities among the selected parameters, as shown in table \ref{Table 3}. In general, model parameters with lower sensitivities can suggest broader sample distributions since their variations make less changes in model outcomes.

\subsection{Uncertainty Propagation from the Model Parameters to the Model Outcomes}
\label{4-2}

As mentioned earlier, only the final results of the model and the corresponding uncertainties are considered important in the context of the robust design. The reason is that such information can help the designers to reduce the influenceof the system performance by the existing uncertainties in the outcomes, regardless of how many variables and models have been involved. Therefore, an appropriate UP technique is essential to propagate uncertainties from the model parameters/input variables to the outputs with a relatively high efficiency and a minimum or at least quantifiable loss of information \cite{ghanem_propagation_1999}. Among different UP methods, first/second order second moment (FOSM/SOSM) approaches are probably the most common approximation procedures in engineering \cite{putko_approach_2001}. In this work, FOSM approach has been used to perform non-linear propagation of the variance-covariance matrix of the model parameters.

Let $\theta$ and $\bar \theta$ vectors denote the optimal MCMC samples for the parameters (after removal of the burn-in period) and their mean values, respectively, and $M(\theta)$ denote the model objective (output) function. If the objective function is estimated by a first order Taylor expansion at $\bar \theta$: 

\begin{equation}
\label{eq 39}
M(\theta) \approx M(\bar \theta) + \sum_{i=1}^{N} {\frac{\partial M(\theta)}{\partial {\theta_i}} (\theta_i-\bar \theta_i)}
\end{equation}

The expected values associated with first and second moment of the above first order Taylor expansion approximates the mean value ($\bar M$) and variance ($\sigma_M^2$) of the objective function $M$ \cite{putko_approach_2001} as follows:

\begin{equation}
\label{eq 40}
\bar M=E[M(\theta)] \approx M(\bar \theta)
\end{equation}

\begin{equation}
\label{eq 41}
\sigma_M^2=E[(M(\theta)-\bar M)^2] \approx \sum_{i=1}^{N} {{|\frac{\partial M(\theta)}{\partial {\theta_i}}|}^2 \sigma_i^2}+\sum_{i=1}^{N}{\sum_{j \neq i}^{N} \frac{\partial M(\theta)}{\partial {\theta_i}} \frac{\partial M(\theta)}{\partial {\theta_j}} \sigma_{ij}}
\end{equation}

where $\sigma_i^2$ and $\sigma_{ij}$ are the variance of the parameter $\theta_i$ and the covariance between the parameters $\theta_i$ and $\theta_j$, respectively. The approximations in equation \ref{eq 40} and \ref{eq 41} has been derived thoroughly by Kriegesmann \cite{kriegesmann_probabilistic_2012}. It is worth noting that equation \ref{eq 41} can also be expressed in terms of the variance-covariance matrix of the model parameters ($V$) \cite{tellinghuisen_statistical_2001-1} as follows:

\begin{equation}
\label{eq 42}
\sigma_M^2 \approx g^T V g
\end{equation}

where g is a column vector which consists of all partial derivative elements, i.e., $g=(\frac{\partial M(\theta)}{\partial {\theta_1}},...,\frac{\partial M(\theta)}{\partial {\theta_N}})^T$.

\begin{figure}[htp]
\centering
\begin{minipage}{0.68\textwidth}
  \centering
  \includegraphics[width=1\linewidth]{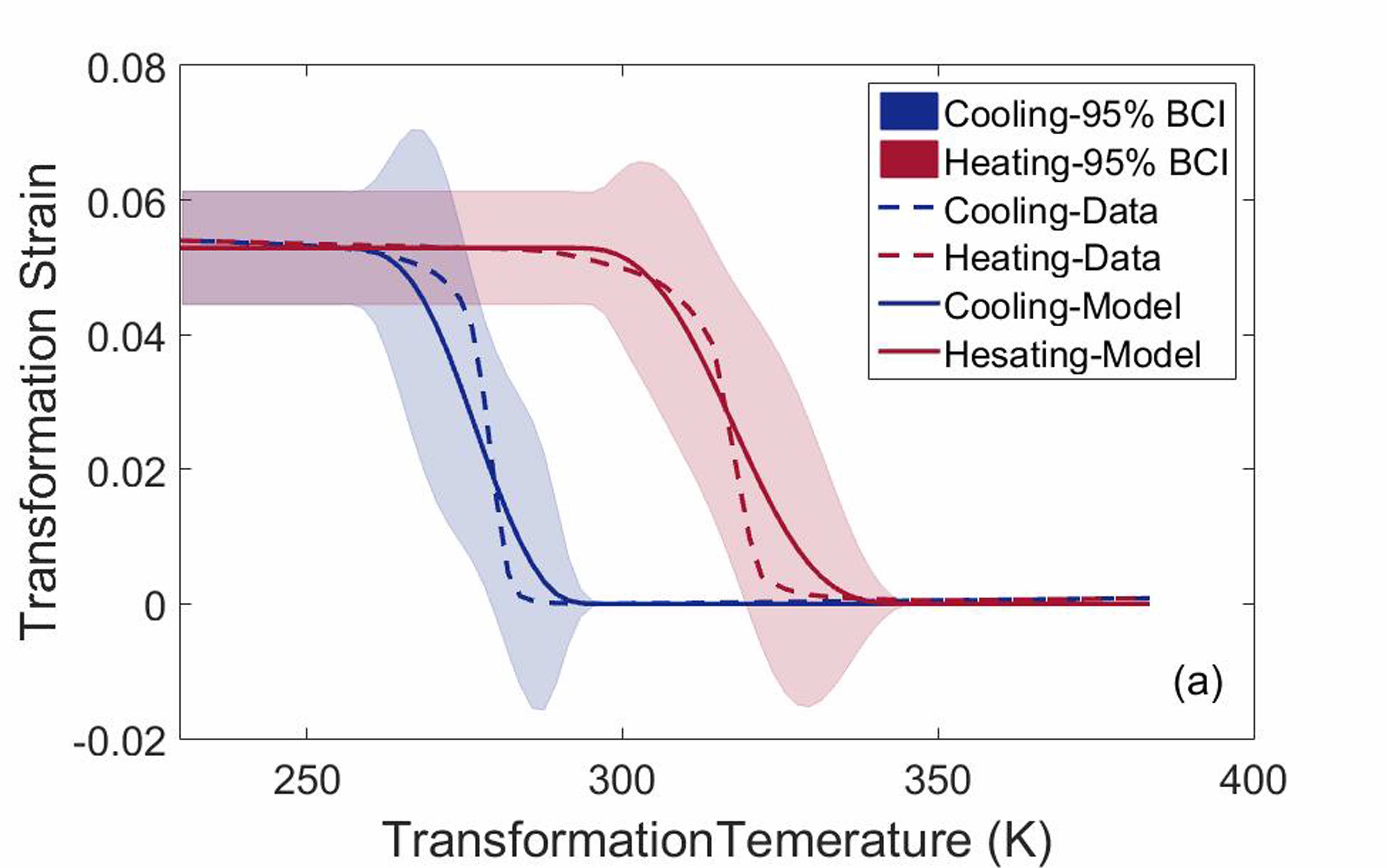}
  \label{fig 4-a}
\end{minipage}
\begin{minipage}{0.68\textwidth}
  \centering
  \includegraphics[width=1\linewidth]{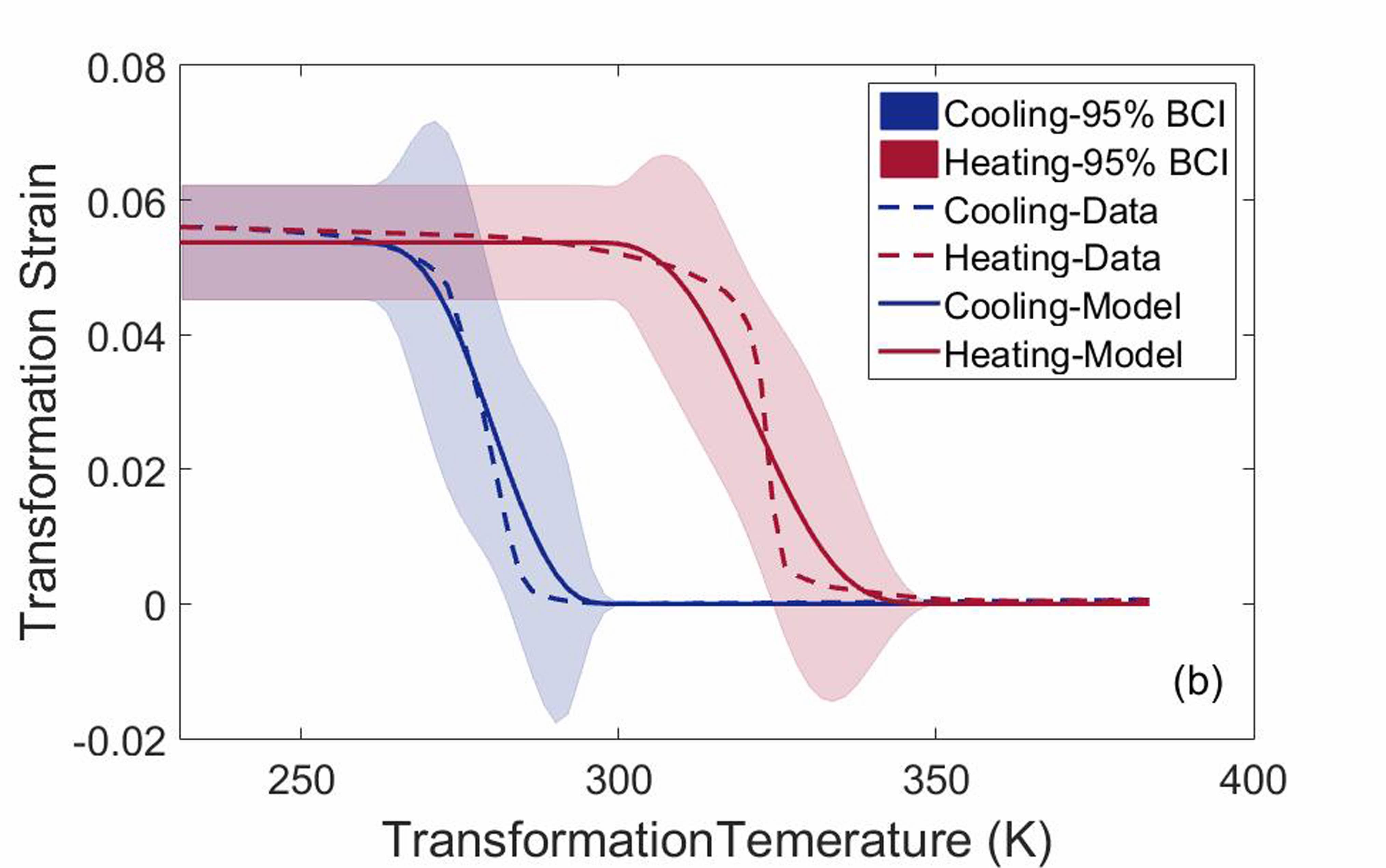}
\label{fig 4-b}
\end{minipage}
\begin{minipage}{0.68\textwidth}
  \centering
  \includegraphics[width=1\linewidth]{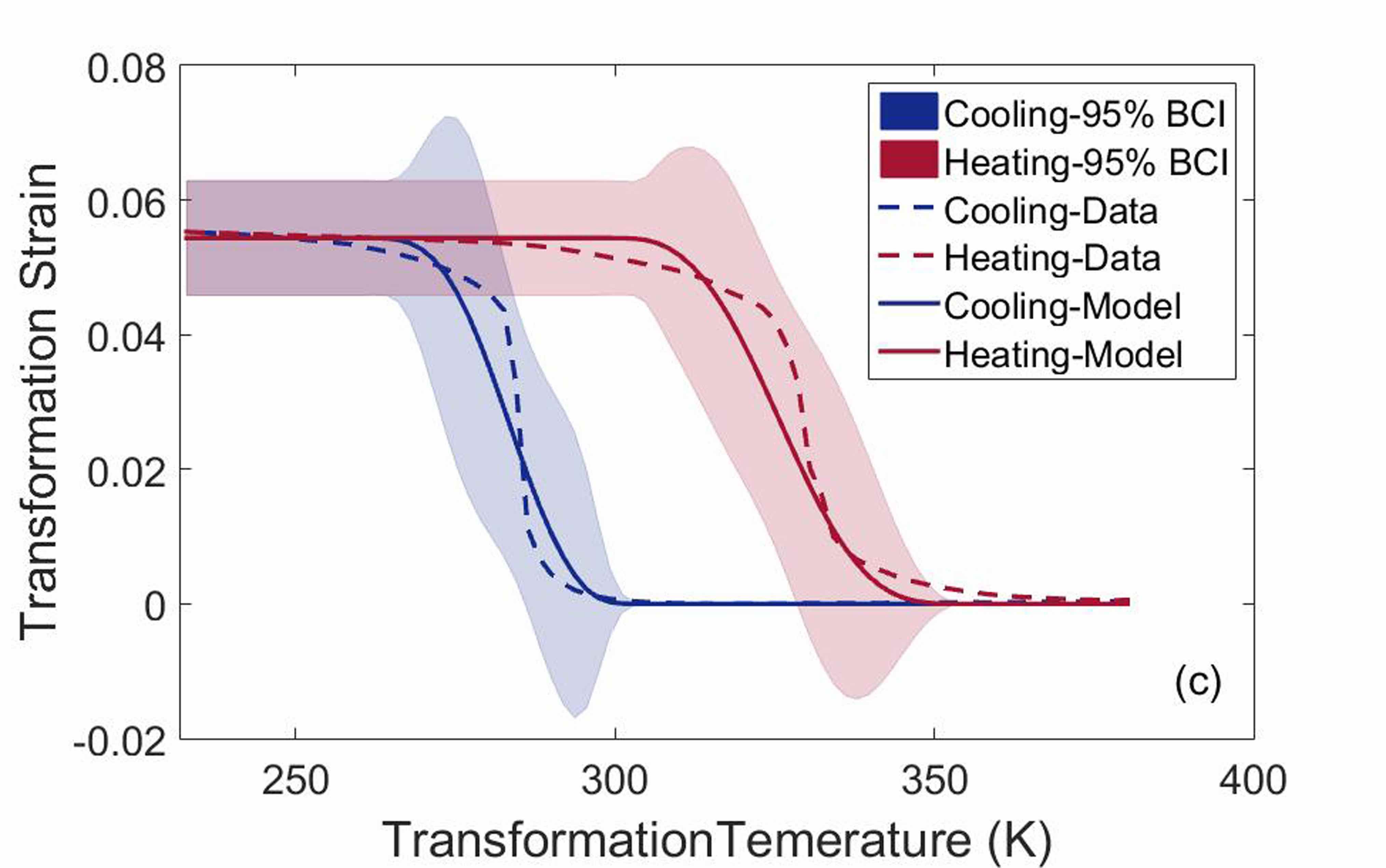}
\label{fig 4-c}
\end{minipage}
\caption{95\% BCIs for the model hysteresis curves obtained from the FOSM approach for different isobaric conditions, a) 100, b) 150, and c) 200 MPa, besides their corresponding experimental counterparts.}
\label{fig 4}
\end{figure}

The uncertainties of parameters reported in table \ref{Table 5} have been propagated to the uncertainty of the transformation strain along the hysteresis curves by applying the variance-covariance matrix obtained from the optimal MCMC samples (without burn-in period) in the above-mentioned FOSM approach. The results have been shown in figure \ref{fig 4} for different isobaric conditions. In these figures, the blue/red solid and dashed lines correspond to the model results during cooling/heating process (forward/reverse martensitic transformation) and their experimental counterparts, respectively. In addition, the blue/red shaded regions indicate 95\% Bayesian confidence interval (95\% BCI) for model results during cooling/heating. It should be noted that the interval is equivalent to $\varepsilon^t(T) \pm 2\sigma_{\varepsilon^t}$, where $\varepsilon^t(T)$ and $\sigma_{\varepsilon^t}$ can be calculated through equations \ref{eq 40} and \ref{eq 41}, respectively. 

As can be observed, there are generally good agreements between the model and experimental hysteresis loops, or at least the experimental data are located inside the 95\% BCIs. Although the given thermo-mechanical model cannot exactly predict the slope of the curves during forward and reverse transformation, there are no discrepancies between model results and experimental data for the maximum transformation strain ($H^{cur}$) in each isobaric condition. The difference in the curves' slopes can be attributed to the slight effects of other parameters that are not considered in the calibration, the missing physics in the model, and the uncertainties in the experimental data all together.

\begin{figure}[htp]
\centering
\begin{minipage}{0.68\textwidth}
  \centering
  \includegraphics[width=1\linewidth]{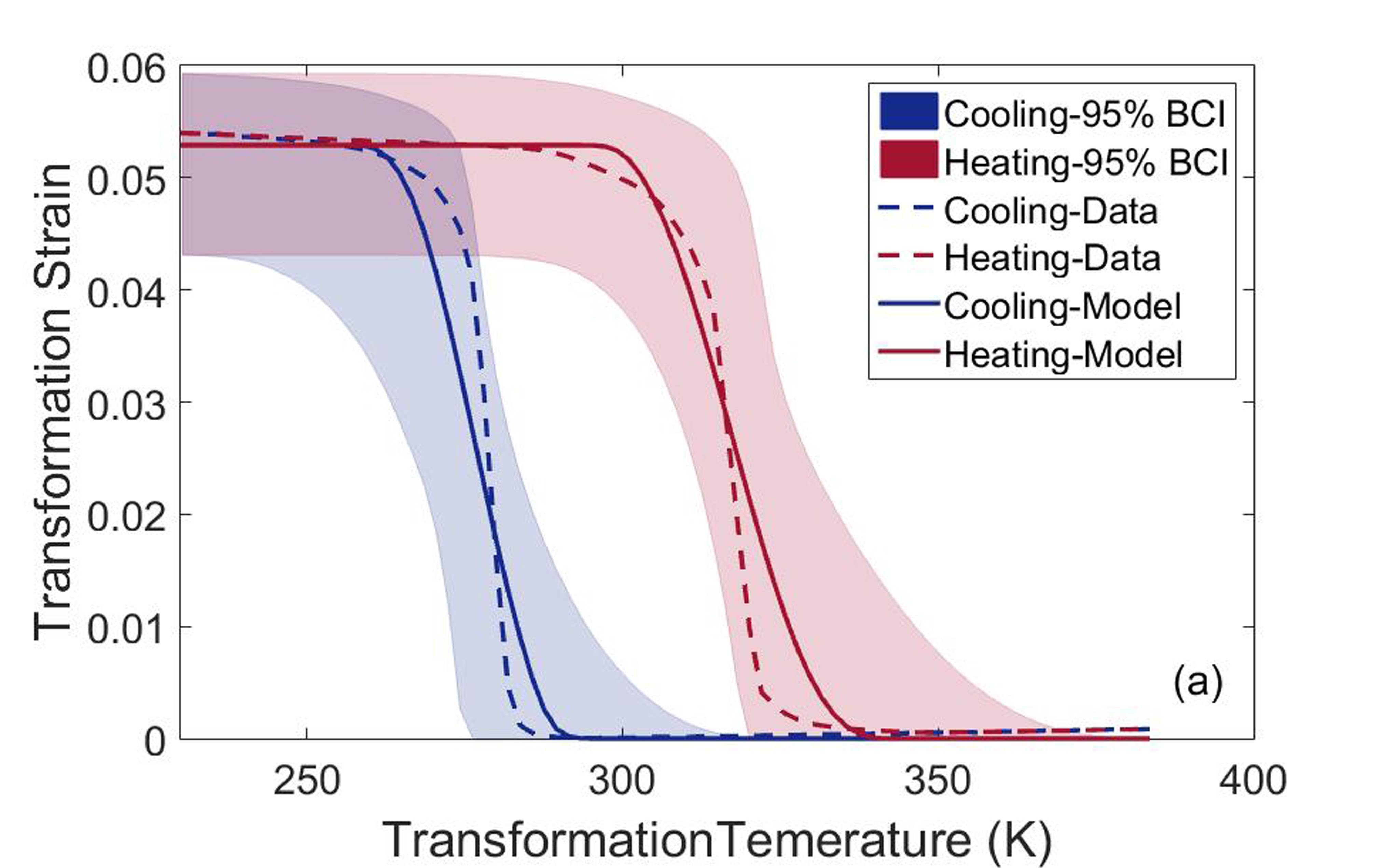}
  \label{fig 5-a}
\end{minipage}
\begin{minipage}{0.68\textwidth}
  \centering
  \includegraphics[width=1\linewidth]{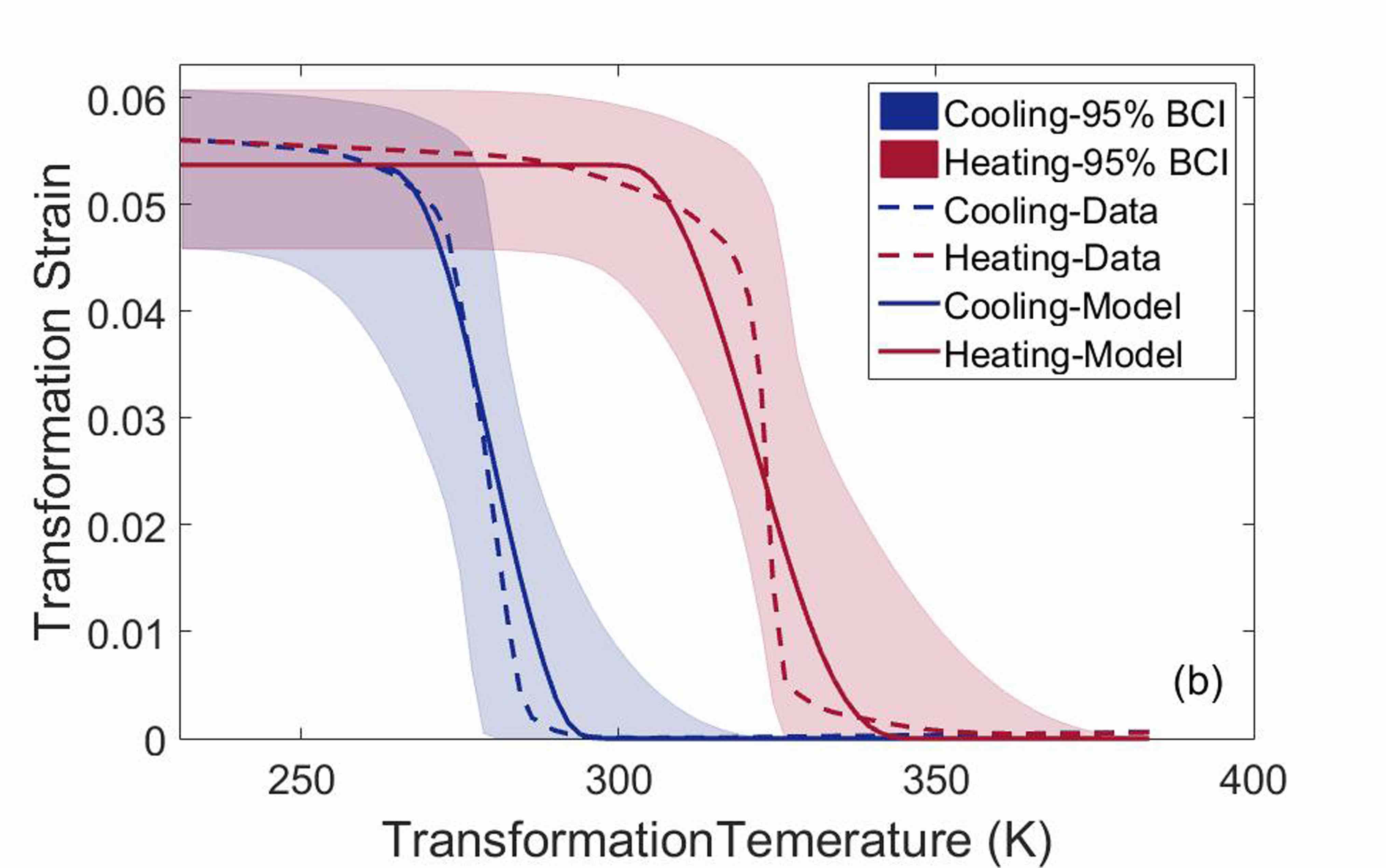}
\label{fig 5-b}
\end{minipage}
\begin{minipage}{0.68\textwidth}
  \centering
  \includegraphics[width=1\linewidth]{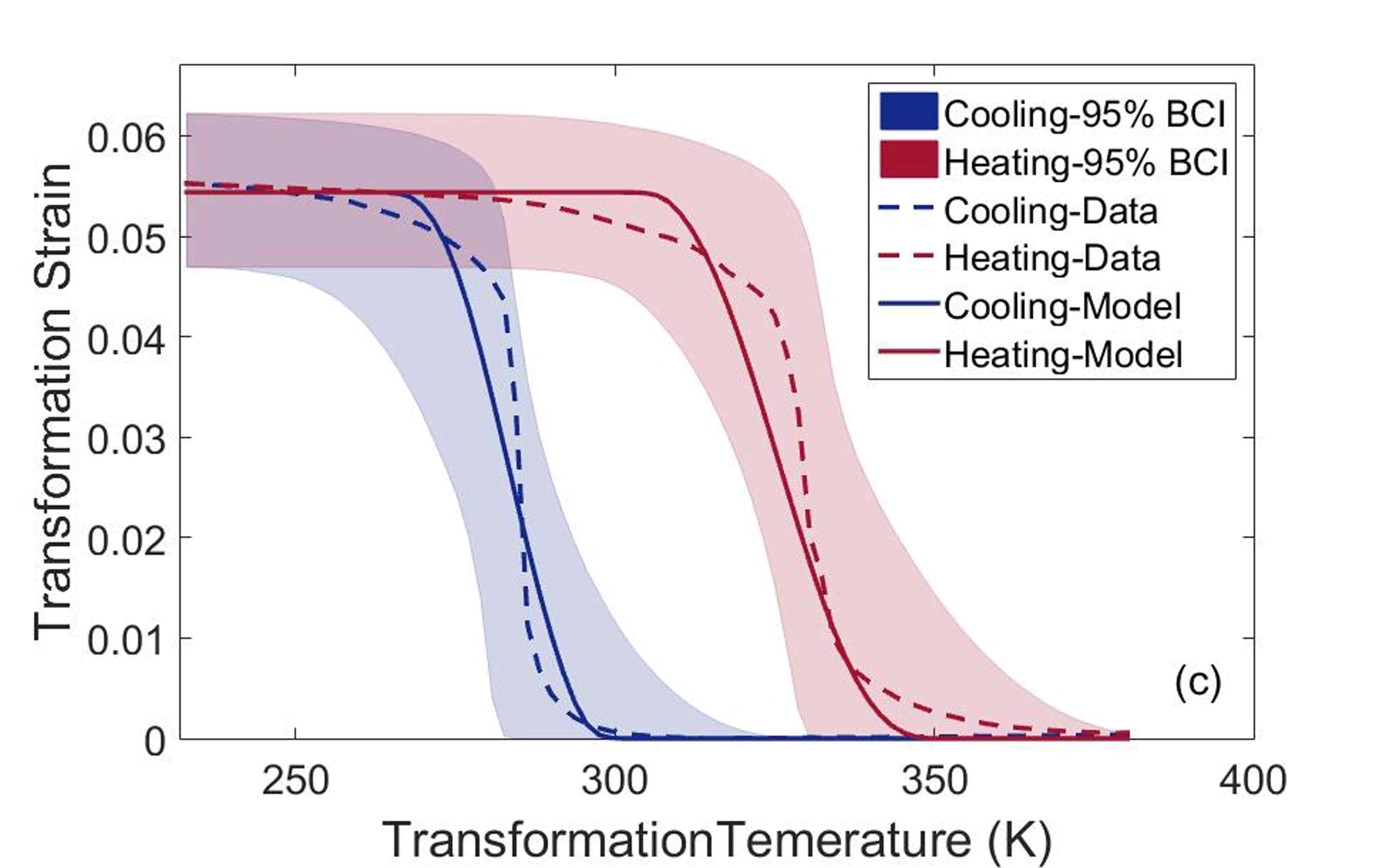}
\label{fig 5-c}
\end{minipage}
\caption{95\% BCIs for the model hysteresis curves obtained from the direct UP approach for different isobaric conditions, a) 100, b) 150, and c) 200 MPa, besides their corresponding experimental counterparts.}
\label{fig 5}
\end{figure}

Another important feature in theses graphs are the unrealistic humps around the transformation temperatures, which can be related to the deficiency of the partial derivatives in the applied UP approach (\ref{eq 41}) at where the change suddenly. For this reason, a direct UP has been performed using the model forward analysis of optimal MCMC samples to find more rational regions for 95\% BCIs. After running the model for all the optimal samples, 2.5\% of the resulting hysteresis curves have been eliminated from each one of the top and down margins to obtain the mentioned intervals. As shown in figure \ref{fig 5}, this approach yields smoother and more precise boundaries for 95\% BCIs with no humps. Although the direct UP approach is more expensive than the FOSM method, it is required to be used in this case to provide more realistic and precise results for 95\% BCIs which are very important in robust design.

\section{MCMC application in decision making for experimental design}
\label{5}

In this section, The goal is to respond the experimentalists' question about which experimental design can gain more knowledge about the system?---performing experiments at the same conditions or at different conditions in design input space. For this purpose, it is assumed that the probabilistically calibrated model in section \ref{4-1} can yield the ground truth and its natural uncertainty bounds at any design condition (i.e., any isobaric condition). According to this assumption, two different synthetic experimental sets are designed through sampling the hysteresis curves from the model uncertainty bounds obtained at any isobaric condition of interest. The first set contains three random samples (replicas) from the same experimental condition (i.e., $\sigma$ = 150 MPa), while the second set considers three random samples from three different conditions (i.e., $\sigma$ = 175, 250, and 300 MPa). After the generation of synthetic data sets, the calibrated results obtained in Table \ref{Table 5} for the influential parameters are considered as the parameter priors which are separately updated against each set of synthetic data using the MCMC technique in a sequential way of data training. In this way of training, the parameter posterior distribution obtained after each training is considered as the prior distribution for the next training. The relative entropy (Kullback–Leibler (K-L) divergence) is calculated to find how diverged the probability distributions are relatively, i.e., what is the distance between the prior and final posterior probability distributions of the parameters after the three sequential MCMC calibrations in each case. The calculated K-L divergence can be introduced as a comparison measure for the amount of information gained from each synthetic experimental sets. Generally, K-L divergence for continuous probability distributions P and Q is defined as:

\begin{equation}
\label{eq 43}
D_{KL}(P||Q) = \int_{-\infty}^{+\infty} p(x) log(\frac{p(x)}{q(x)}) dx
\end{equation}

where $p(x)$ and $q(x)$ are the densities of $P$ and $Q$ at any given random point $x$, respectively. In the case of multivariate Gaussian distribution, the integration in Equation \ref{eq 43} can be obtained as follows:

\begin{equation}
\label{eq 44}
D_{KL}(\mathcal{N}_1||\mathcal{N}_2) = \frac{1}{2} \Big[ln(\frac{|\Sigma_2|}{|\Sigma_1|})-d+tr(\Sigma_2^{-1}\Sigma_1)+(\mu_2-\mu_1)^T\Sigma_2^{-1}(\mu_2-\mu_1) \Big]
\end{equation}

The equivalent normal distributions obtained for parameter prior and posterior distributions are compared for each set of experimental design using Equation \ref{eq 44}. The K-L divergence values for experimental sets 1 and 2 are calculated around 3.8 and 4.4, respectively. The higher value of the K-L divergence for experimental set 2 indicates a greater difference between prior and final posterior distributions in this case, which implies that the consideration of different conditions in experimental design can gain more information about the system rather than experimental replicas in the same condition. 

\section{Summary and Conclusion}
\label{6}

In this work, the calibration and UQ of the sensitive parameters in a thermo-mechanical model have been performed against three different isobaric experimental data simultaneously using a constrained MCMC-MH algorithm in the context of Bayesian statistics. Eight sensitive parameters have been found before the calibration using a DOE approach which includes CFD and N-way ANOVA. After MCMC sampling of these eight parameters, their convergence have been checked by the joint frequency distributions and the cumulative mean plots, which have also been utilized to identify the qualitative correlation of each pair parameters and the burn-in period, respectively. In addition, the linear correlation between each two parameters has been calculated through Pearson coefficient. These coefficients suggest that most pair parameters are linearly uncorrelated, although a few of them show weak or moderate linear correlations. 

After removal of the samples belonging to the burn-in period, skewed Gaussian distributions have been obtained for the marginal posterior frequency distributions of the model parameters, which are truncated in the range defined for each parameter. Among the selected parameters, the parameters with the lowest sensitivities show flatter distributions due to their less influences on the model outcome. Moreover, it should be noted that the mean and square root of the diagonal elements in the variance-covariance matrix of the remaining optimal samples after the elimination of the burn-in period provide the plausible optimal values and uncertainties of the selected parameters, respectively. 

At the end, the parameters' uncertainties have been propagated to the uncertainties of the transformation strain along the hysteresis curves using two various UP techniques in order to find 95\% BCIs during cooling/heating process. In this regard, FOSM approach results in some unrealistic humps around the transformation temperatures. For this reason, a direct UP technique has been performed through the forward analysis of optimal MCMC samples in order to achieve more rational and precise uncertainty bands that are very important in the context of robust design. The hysteresis curves obtained from the mean values of the parameters are in good agreements with their experimental counterparts for different given isobaric conditions, or at least it can be stated that the experimental data are situated in 95\% BCIs. Although the slopes of the forward and reverse transformation curves are not predicted exactly, the maximum transformation strain associated with each isobaric condition is very close to its corresponding data. 

In this work, It has been also shown that the Markov Chain Monte Carlo approach can be applied for decision making in experimental design. Assuming the calibrated model can predict the ground truth and the natural uncertainties in different experimental conditions, two synthetic experimental sets have been sampled from the model response uncertainty bounds obtained at the same and different isobaric conditions, respectively. After sequential MCMC updates of parameters' posterior distributions with each set, their information gains are compared together through the calculation of K-L divergence for each case which indicate the difference between prior and final posterior distributions of the model parameters. A higher information gain has been obtained from the experimental set 2 that are sampled from different experimental conditions (different isobaric conditions).

Generally, this work introduces a complete framework for UQ of model parameters and subsequent UP from model parameters to outputs as a guideline for material design.

\section*{References}
\smallskip
\bibliographystyle{iopart-num}
\bibliography{Ref}

\providecommand{\newblock}{}
\begin{thebibliography}{10}
\expandafter\ifx\csname url\endcsname\relax
  \def\url#1{{\tt #1}}\fi
\expandafter\ifx\csname urlprefix\endcsname\relax\def\urlprefix{URL }\fi
\providecommand{\eprint}[2][]{\url{#2}}

\bibitem{kang_thermomechanical_2017}
Kang G and Kan Q 2017 Thermomechanical cyclic deformation of shape-memory
  alloys {\em Cyclic Plasticity of Engineering Materials\/} (John Wiley \&
  Sons, Ltd) pp 405--530

\bibitem{kan_constitutive_2010}
Kan Q and Kang G 2010 {\em Int. J. Plast.\/} {\bf 26} 441--465

\bibitem{yu_physical_2017}
Yu C, Kang G, Kan Q and Xu X 2017 {\em Mech. Mater.\/} {\bf 112} 1--17

\bibitem{yu_macroscopic_2017}
Yu C, Kang G and Kan Q 2017 {\em Acta Mech. Sin.\/} {\bf 33} 619--634

\bibitem{bhaumik_nickeltitanium_2014}
Bhaumik S~K, Ramaiah K~V and Saikrishna C~N 2014 {Nickel}-titanium shape memory
  alloy wires for thermal actuators {\em Micro and Smart Devices and Systems\/}
  Springer Tracts in Mechanical Engineering (Springer, New Delhi) pp 181--198

\bibitem{dilibal_nickeltitanium_2017}
Dilibal S, Sahin H, Dursun E and Engeberg E~D 2017 {\em Electr. Eng.\/} {\bf
  99} 923--930

\bibitem{baxevanis_micromechanics_2014}
Baxevanis T, Cox A and Lagoudas D~C 2014 {\em Acta Mech.\/} {\bf 225}
  1167--1185

\bibitem{patoor_shape_2006}
Patoor E, Lagoudas D~C, Entchev P~B, Brinson L~C and Gao X 2006 {\em Mech.
  Mater.\/} {\bf 38} 391--429

\bibitem{povoden-karadeniz_thermodynamics_2013-1}
Povoden-Karadeniz E, Cirstea D~C, Lang P, Wojcik T and Kozeschnik E 2013 {\em
  Calphad\/} {\bf 41} 128--139

\bibitem{frost_thermomechanical_2010}
Frost M, Sedlák P, Sippola M and Šittner P 2010 {\em Smart Mater. Struct.\/}
  {\bf 19} 094010

\bibitem{johnson_inverse_2016-1}
Johnson L and Arróyave R 2016 {\em Mater. Des.\/} {\bf 107} 7--17

\bibitem{tapia_bayesian_2017}
Tapia G, Johnson L, Franco B, Karayagiz K, Ma J, Arroyave R, Karaman I and
  Elwany A 2017 {\em J. Manuf. Sci. Eng\/} {\bf 139} 071002--071002--13

\bibitem{steinbach_phase-field_2011}
Steinbach I and Shchyglo O 2011 {\em Curr. Opin. Solid State Mater. Sci.\/}
  {\bf 15} 87--92

\bibitem{ke_modeling_2012-1}
Ke C~b, Cao S~s, Ma X and Zhang X~p 2012 {\em Trans. Nonferrous Met. Soc.
  China\/} {\bf 22} 2578--2585

\bibitem{ke_phase_2015-1}
Ke C~B, Cao S~S and Zhang X~P 2015 {\em Modelling Simul. Mater. Sci. Eng.\/}
  {\bf 23} 055008

\bibitem{yu_macroscopic_2017-1}
Yu C, Kang G and Kan Q 2017 {\em Acta Mech. Sin.\/} {\bf 33} 619--634

\bibitem{yu_physical_2017-1}
Yu C, Kang G, Kan Q and Xu X 2017 {\em Mech. Mater\/} {\bf 112} 1--17

\bibitem{haller_thermomechanical_2016}
Haller L, Nedjar B, Moumni Z, Vedinaş I and Trană E 2016 {\em Continuum Mech.
  Thermodyn.\/} {\bf 28} 957--975

\bibitem{kan_constitutive_2010-1}
Kan Q and Kang G 2010 {\em Int. J. Plast.\/} {\bf 26} 441--465

\bibitem{jiang_propagation_2017}
Jiang D, Kyriakides S and Landis C~M 2017 {\em Extreme Mech. Lett.\/} {\bf 15}
  113--121

\bibitem{lagoudas_constitutive_2012}
Lagoudas D, Hartl D, Chemisky Y, Machado L and Popov P 2012 {\em Int. J.
  Plast.\/} {\bf 32-33} 155--183

\bibitem{saleeb_multi-axial_2011}
Saleeb A~F, Padula S~A and Kumar A 2011 {\em Int. J. Plast.\/} {\bf 27}
  655--687

\bibitem{bodaghi_robust_2014}
Bodaghi M, Damanpack A~R, Aghdam M~M and Shakeri M 2014 {\em Int. J. Eng.
  Sci.\/} {\bf 82} 1--21

\bibitem{song_closed-form_2015}
Song Z and Dai H~H 2015 {\em Int. J. Eng. Sci.\/} {\bf 88} 40--63

\bibitem{zhu_determining_2016}
Zhu P, Feng P, Sun Q~P, Wang J and Dai H~H 2016 {\em Int. J. Plast.\/} {\bf 85}
  52--76

\bibitem{mirzaeifar_micromechanical_2013}
Mirzaeifar R, {DesRoches} R, Yavari A and Gall K 2013 {\em Acta Mater.\/} {\bf
  61} 4542--4558

\bibitem{yu_micromechanical_2013}
Yu C, Kang G, Kan Q and Song D 2013 {\em Int. J. Plast.\/} {\bf 44} 161--191

\bibitem{yu_micromechanical_2015}
Yu C, Kang G and Kan Q 2015 {\em J. Mech. Phys. Solids\/} {\bf 82} 97--136

\bibitem{heinen_micromechanical_2012}
Heinen R and Miro S 2012 {\em Computer Methods in Applied Mechanics and
  Engineering\/} {\bf 229-232} 44--55

\bibitem{lagoudas_shape_2006}
Lagoudas D~C, Entchev P~B, Popov P, Patoor E, Brinson L~C and Gao X 2006 {\em
  Mech. Mater.\/} {\bf 38} 430--462

\bibitem{choi_inductive_2008-1}
Choi H~J, Mcdowell D~L, Allen J~K and Mistree F 2008 {\em Eng. Optimiz.\/} {\bf
  40} 287--307

\bibitem{hill_mary_c._methods}
{Hill M C} 2000 {\em \normalfont{In} \textit{Building Partnerships}\/}  1--10

\bibitem{Kim_deterministic_2014}
Kim Y~J, Kim K~C, Park C~S and Kim I~H 2014 {\em \normalfont{In} \textit{The
  2nd Asia conference of International Building Performance Simulation
  Association}\/}  28--29

\bibitem{larssen_forecasting_2006}
Larssen T, Huseby R~B, Cosby B~J, Høst G, Høgåsen T and Aldrin M 2006 {\em
  Environ. Sci. Technol.\/} {\bf 40} 7841--7847

\bibitem{jackman_estimation_2000}
Jackman S 2000 {\em Am. J. Political Sci.\/} {\bf 44} 375--404

\bibitem{olbricht_bayes_1994-1}
Olbricht W, Chatterjee N~D and Miller K 1994 {\em Phys. Chem. Minerals\/} {\bf
  21} 36--49

\bibitem{lynch_introduction_2007-1}
Lynch S~M 2007 {\em Introduction to Applied Bayesian Statistics and Estimation
  for Social Scientists\/} (Springer Science \& Business Media)

\bibitem{saltelli_variance_2010}
Saltelli A, Annoni P, Azzini I, Campolongo F, Ratto M and Tarantola S 2010 {\em
  Comput. Phys. Commun.\/} {\bf 181} 259--270

\bibitem{fisher_design_1966}
Fisher Ronald~Aylmer S 1966 {\em The design of experiments\/} 8th ed (Edinburgh
  ; London : Oliver \& Boyd)

\bibitem{ghanem_propagation_1999}
Ghanem R and Red-Horse J 1999 {\em Physica D: Nonlinear Phenomena\/} {\bf 133}
  137--144

\bibitem{paglietti_mathematical_1977}
Paglietti A 1977 {\em Ann. Inst. Henri Poincaré, Section (A): Physique
  Theorique\/} {\bf 27} 207--219

\bibitem{coleman_thermodynamics_1963}
Coleman B~D and Noll W 1963 {\em Arch. Rational Mech. Anal.\/} {\bf 13}
  167--178

\bibitem{tschopp_quantifying_2017}
Tschopp M~A and Hernandez-Rivera E 2017 Quantifying similarity and distance
  measures for vector-based datasets: Histograms, signals, and probability
  distribution functions \unskip\space {ARL}-{TN}-0810 {US} Army Research
  Laboratory Aberdeen Proving Ground United States, {US} Army Research
  Laboratory Aberdeen Proving Ground United States

\bibitem{billera_geometric_2001}
Billera L~J and Diaconis P 2001 {\em Statistical Science\/} {\bf 16} 335--339

\bibitem{Gelman_Bayesian_2014}
Gelman A, Carlin J~B, Stern H~S, Dunson D~B, Vehtari A and Rubin D~B 2014 {\em
  Bayesian Data Analysis\/} vol~2 (Boca Raton, FL: CRC press)

\bibitem{haario_dram:_2006}
Haario H, Laine M, Mira A and Saksman E 2006 {\em Stat. Comput.\/} {\bf 16}
  339--354

\bibitem{haario_adaptive_2001}
Haario H, Saksman E and Tamminen J 2001 {\em Bernoulli\/} {\bf 7} 223--242

\bibitem{putko_approach_2001}
Putko M~M, Taylor Arthur~C I, Newman P~A and Green L~L 2001 {\em J. Fluids
  Eng.\/} {\bf 124} 60--69

\bibitem{kriegesmann_probabilistic_2012}
Kriegesmann B 2012 {\em Probabilistic Design of Thin-walled Fiber Composite
  Structures\/} ({ISD})

\bibitem{tellinghuisen_statistical_2001-1}
Tellinghuisen J 2001 {\em J. Phys. Chem. A\/} {\bf 105} 3917--3921

\bibitem{montgomery_design_2017}
Montgomery D~C 2017 {\em Design and Analysis of Experiments\/} (John Wiley \&
  Sons)

\end{thebibliography}

\section*{Appendix A: Two-way ANOVA}

In the case of two-way ANOVA where there are $L$ observations and two multi-level factors/parameters/variables $A$ and $B$, system response can be defined as a Means model as follows \cite{montgomery_design_2017}:

\begin{equation}
\label{eq 24}
y_{ijl}=\mu_{ij}+\epsilon_{ijl}=\mu+\tau_{i}+\beta_{j}+(\tau \beta)_{ij}+\epsilon_{ijl} \ \ , \ \ \Bigg \{\begin{array}{l}
            i=1,...,I \\
            j=1,...,J \\
            l=1,...,L
        \end{array}
\end{equation}

where $I$ and $J$ are the number of levels associated with factors/parameters/variables A and B, and $\tau_{i}$, $\beta_{j}$, and $(\tau \beta)_{ij}$ denote the effects of the level $i$ of factor $A$, level $j$ of factor $B$, and their interaction, respectively. $\epsilon$ is the residual or the random error which is considered as an independent normally distributed function with a fixed variance, $\mathcal{N}(0,\sigma^2)$. In this context, the following hypotheses are statistically evaluated to find the most sensitive factors/parameters/variables and interactions in the system \cite{montgomery_design_2017}:

\begin{eqnarray}
H_0: \ \ \tau_1=\tau_2=...=\tau_I=0 \nonumber \\
H_1: \ \ \textnormal{at least one } \tau_i \neq 0 \label{eq 25}\\
\nonumber \\
H_0: \ \ \beta_1=\beta_2=...=\beta_J=0 \nonumber \\
H_1: \ \ \textnormal{at least one } \beta_j \neq 0 \label{eq 26}\\
\nonumber \\
H_0: \ \ (\tau \beta)_{ij}=0 \ \ \textnormal{for all $i$ and $j$} \nonumber \\
H_1: \ \ \textnormal{at least one } (\tau \beta)_{ij} \neq 0 \label{eq 27}
\end{eqnarray}

In each case, $H_0$ and $H_1$ are null and alternative hypothesis, respectively. It should be noted that a component (either a factor/parameter/variable or an interaction) is identified sensitive when the corresponding null hypothesis is rejected after hypothesis testing. In this regard, a variance partitioning into the components of the system is required; however, this partitioning can be performed on the total sum of squares instead due to the proportionality between total variance and total sum of squares. 

Let $y_{i..}$, $y_{.j.}$, $y_{ij.}$, and $y_{...}$ denote the sum of all the factorial design responses which includes the effects of the level $i$ of the factor/parameter/variable $A$, the level $j$ of the factor/parameter/variable $B$, the combination of the level $i$ for $A$ and the level $j$ for $B$, and all the level-factor/parameter/variable combinations, respectively. Moreover, let $\bar{y}_{i..}$, $\bar{y}_{.j.}$, $\bar{y}_{ij.}$, and $\bar{y}_{...}$ denote the corresponding averages as follows \cite{montgomery_design_2017}:

\begin{eqnarray}
y_{i..}=\displaystyle\sum_{j=1}^{J} \displaystyle\sum_{l=1}^{L} y_{ijl} \ \ , \ \ \bar{y}_{i..}=\frac{y_{i..}}{JL} \label{eq 28} \\
y_{.j.}=\displaystyle\sum_{i=1}^{I} \displaystyle\sum_{l=1}^{L} y_{ijl} \ \ , \ \ \bar{y}_{.j.}=\frac{y_{.j.}}{IL} \label{eq 29} \\
y_{ij.}=\displaystyle\sum_{l=1}^{L} y_{ijl} \ \ , \ \ \bar{y}_{ij.}=\frac{y_{ij.}}{L} \label{eq 30} \\
y_{...}=\displaystyle\sum_{i=1}^{I} \displaystyle\sum_{j=1}^{J} \displaystyle\sum_{l=1}^{L} y_{ijl} \ \ , \ \ \bar{y}_{...}=\frac{y_{...}}{IJL} \label{eq 31}
\end{eqnarray}

Total sum of squares ($SS_T$) can be decomposed to the sum of squares of the system components, as follows:

\begin{eqnarray}
\label{eq 32}
\hspace*{-1.5cm} SS_T=\displaystyle\sum_{i=1}^{I} \displaystyle\sum_{j=1}^{J} \displaystyle\sum_{l=1}^{L} (y_{ijl}-\bar{y}_{...})^2&=\displaystyle\sum_{i=1}^{I} \displaystyle\sum_{j=1}^{J} \displaystyle\sum_{l=1}^{L} \Big[(\bar{y}_{i..}-\bar{y}_{...})+(\bar{y}_{.j.}-\bar{y}_{...}) \nonumber \\
& \ \ +(\bar{y}_{ij.}-\bar{y}_{i..}-\bar{y}_{.j.}+\bar{y}_{...})+({y}_{ijk}-\bar{y}_{ij.}) \Big]^2 \nonumber \\
&= JL \displaystyle\sum_{i=1}^{I} (\bar{y}_{i..}-\bar{y}_{...})^2 + IL \displaystyle\sum_{j=1}^{J} (\bar{y}_{.j.}-\bar{y}_{...})^2 \nonumber \\
& \ \ +L \displaystyle\sum_{i=1}^{I} \displaystyle\sum_{j=1}^{J} (\bar{y}_{ij.}-\bar{y}_{i..}-\bar{y}_{.j.}+\bar{y}_{...})^2 \nonumber \\
& \ \ +\displaystyle\sum_{i=1}^{I} \displaystyle\sum_{j=1}^{J} \displaystyle\sum_{l=1}^{L} ({y}_{ijl}-\bar{y}_{ij.})^2 \nonumber \\
&= SS_A + SS_B + SS_{AB} + SS_E
\end{eqnarray}

where $SS_A$, $SS_B$, $SS_{AB}$, $SS_E$ are the sum of squares associated with $A$, $B$, $A$-$B$ interaction, and the random error, respectively. It is worth noting that $SS_E$ can be obtained by the subtraction of $SS_A$, $SS_B$, and $SS_{AB}$ from $SS_T$. In equation \ref{eq 32}, the first equality results from the addition and subtraction of the same terms on the right side, and the second equality is a consequence of the polynomial expansion of the function where the six cross products becomes zero. If these sum of squares are divided by their corresponding degrees of freedom, the mean squares ($MS$) are obtained. The expected values of these mean squares when the levels are random variables can provide relationships to estimate the total deviations of all the levels associated with $A$, $B$, and $A$-$B$ interaction, as follows:

\begin{eqnarray}
E(MS_A)=E\Big(\frac{SS_A}{I-1} \Big)=\sigma^2 + \frac{JL \displaystyle\sum_{i=1}^{I} \tau_i^2}{I-1} \label{eq 33} \\
E(MS_B)=E\Big(\frac{SS_B}{J-1} \Big)=\sigma^2 + \frac{IL \displaystyle\sum_{j=1}^{J} \beta_j^2}{J-1} \label{eq 34}\\
E(MS_{AB})=E\Big(\frac{SS_{AB}}{(I-1)(J-1)} \Big)=\sigma^2 + \frac{L \displaystyle\sum_{i=1}^{I} \displaystyle\sum_{j=1}^{J} (\tau \beta)_{ij}^2}{(I-1)(J-1)} \label{eq 35} \\
\nonumber \\
E(MS_E)=E\Big(\frac{SS_E}{IJ(L-1)} \Big)=\sigma^2 \label{eq 36}
\end{eqnarray}

It should be noted that the mean square of each factor/parameter/variable or interaction can be used in the above equations in the problems that the levels are fixed. In the case that the null hypothesis about the insensitivity of a factor/parameter/variable or an interaction is true, the total corresponding effects of the component becomes zero; therefore, the corresponding expected value in equation \ref{eq 33}, \ref{eq 34} or/and \ref{eq 35} equals the expected value for the mean squares of the random error, i.e., its variance ($\sigma^2$). On the other hand, if there are any effects from the system components, the corresponding expected values are greater than $\sigma^2$. It is clear that bigger ratio of the expected value of a component to $\sigma^2$ is equivalent to more sensitivity of the component. This is also true for the ratio of the corresponding mean squares, which is called $F$-ratio. However, a certain criterion is required to fully reject the null hypotheses for the identification of the sensitive components in the system. For this reason, a $F$-distribution is created from anyone of the null hypotheses based on the numerator and denominator degrees of freedom of $F$-ratio to calculate the corresponding p-value. Generally, the area underneath the F-distribution confined between the F-ratio value and infinity is introduced as the p-value. Different significance levels can be set as the criteria to reject the null hypotheses, i.e., 0.01, 0.05, or 0.1. For any system component, a p-value less than the significance level rejects the corresponding null hypothesis, and demonstrates the sensitivity of the given component.

In the case that there is just one observation/response for each level-factor/parameter/variable combinations (generally for the models and simulations where their response is deterministic such as our thermo-mechanical model), the variance of the random error ($\sigma^2$) can not be determined by the expected mean squares of the random error in equation \ref{eq 36} since $L$ equals $1$. Under this condition, no clear differentiation can be established between the effect of the interaction and the random error in equation \ref{eq 35}. Therefore, the interaction effect in equation \ref{eq 24} must be considered zero in order to be able to estimate $\sigma^2$ using equation \ref{eq 35}; otherwise, the effect of each individual factor/parameter/variable can not be estimated through equations \ref{eq 33} and \ref{eq 34}. ANOVA table for this case can be observed in table \ref{Table 2} \cite{montgomery_design_2017}.

\begin{table}[t]
\caption{\label{Table 2} Two-way ANOVA table for the case of one observation/response per level-factor/parameter/variable combination \cite{montgomery_design_2017}.}
\begin{indented}
\item[]\begin{tabular}{@{}llllll}
\br
Source of & Sum of & Degrees of & Mean & Expected Mean & F- \\
Variation & Squares & Freedoms & Square & Square & Ratio \\
\mr
$A$ & $J \displaystyle\sum_{i=1}^{I} (\bar{y}_{i.}-\bar{y}_{..})^2$ & $I-1$ & $MS_A$ & $\sigma^2 + \frac{J \sum \tau_i^2}{I-1}$ & $\frac{MS_A}{MS_E}$\\
$B$ & $I \displaystyle\sum_{j=1}^{J} (\bar{y}_{.j}-\bar{y}_{..})^2$ & $J-1$ & $MS_B$ & $\sigma^2 + \frac{I \sum \beta_j^2}{J-1}$ & $\frac{MS_B}{MS_E}$ \\
Residual or $AB$ & Subtraction & $(I-1)(J-1)$ & $MS_E$ & $\sigma^2+\frac{\sum \sum (\tau \beta)_{ij}^2}{(I-1)(J-1)}$ \\
Total & $\displaystyle\sum_{i=1}^{I} \displaystyle\sum_{j=1}^{J} (y_{ij} - \bar{y}_{..})^2$ & $IJ-1$ \\
\br
\end{tabular}
\end{indented}
\end{table}

\end{document}